\documentclass[a4paper,useAMS,usenatbib]{mnras}
\pdfoutput=1

\usepackage{graphicx}
\usepackage{setspace}
\usepackage{natbib}
\usepackage{color}
\usepackage{amsmath,amssymb}
\usepackage{times}
\usepackage{hyperref}[]

\newcommand{\bprp}{\ensuremath{\mathrm{BP}-\mathrm{RP}}}

\title[Unresolved stellar companions with {\it Gaia} DR2 astrometry]{Unresolved stellar companions with {\it Gaia} DR2 astrometry}

\author[Belokurov et al]{Vasily
  Belokurov$^{1}$\thanks{E-mail:vasily@ast.cam.ac.uk}, Zephyr Penoyre$^1$, Semyeong Oh$^1$, Giuliano Iorio$^{1,2}$, Simon Hodgkin$^1$,
  \newauthor N. Wyn Evans$^{1}$, Andrew Everall$^1$, Sergey E. Koposov$^{3,1,4}$, Christopher A. Tout$^1$
  \newauthor Robert Izzard$^5$, Cathie J. Clarke$^1$ and Anthony G. A. Brown$^6$\\$^1$Institute of Astronomy, Madingley Rd,
    Cambridge, CB3 0HA\\
$^{2}$Dipartimento di Fisica e Astronomia ”G. Galilei”, Universit\`a di Padova, vicolo dell’Osservatorio 3, 35122 PD, Italy\\
    $^{3}$McWilliams Center for Cosmology, Carnegie Mellon University, 5000 Forbes Ave, 15213, USA\\
    $^{4}$Kavli
 Institute for Cosmology, University of Cambridge, Madingley Road, Cambridge CB3 0HA, UK\\
    $^5$Astrophysics Research Group, University of Surrey, Guildford, Surrey, GU2 7XH\\
$^6$ Leiden Observatory, Leiden University, Niels Bohrweg 2, 2333 CA Leiden, The Netherlands}

\hypersetup{draft}
\begin{document}



\maketitle

\label{firstpage}

\begin{abstract}
For stars with unresolved companions, motions of the centre of light
and that of mass decouple, causing a single-source astrometric model
to perform poorly. We show that such stars can be easily detected with
the reduced $\chi^2$ statistic, or RUWE, provided as part of {\it
  Gaia} DR2. We convert RUWE into the amplitude of the image centroid
wobble, which, if scaled by the source distance, is proportional to
the physical separation between companions (for periods up to several
years). We test this idea on a sample of known spectroscopic binaries
and demonstrate that the amplitude of the centroid perturbation scales
with the binary period and the mass ratio as expected. We apply this
technique to the {\it Gaia} DR2 data and show how the binary fraction
evolves across the Hertzsprung--Russell diagram. The observed
incidence of unresolved companions is high for massive young stars and
drops steadily with stellar mass, reaching its lowest levels for white
dwarfs. We highlight the elevated binary fraction for the nearby Blue
Stragglers and Blue Horizontal Branch stars. We also illustrate how
unresolved hierarchical triples inflate the relative velocity signal
in wide binaries. Finally, we point out a hint of evidence for the
existence of additional companions to the hosts of extrasolar hot
jupiters.

\end{abstract}


\section{Introduction}

A star's path on the sky is often wiggled, but not always due to its
parallax. Unresolved stellar companions induce photocentre wobble
giving us a chance to detect binary systems via astrometry. This was
first demonstrated almost a century ago
\citep[see][]{Reuyl1936,Lippincott1955}. Better still, the motion of
the centre of light can be straightforwardly interpreted, placing
constraints on the properties of the unseen companion
\citep[see][]{vandenKamp1975}. Space-based astrometric missions such
as {\it Hipparcos} \citep[][]{Perryman1997} and {\it Gaia}
\citep[][]{Perryman2001, Prusti2016} have offered a much improved
chance of discovering small wobbles in the stellar motion due to
multiplicity. Inspired by this, the community has understandably
focused on stellar companions that are tricky to observe otherwise
such as exosolar planets
\citep[][]{Lattanzi1997,Sozzetti2001,Casertano2008,Perryman2014} and
dark remnants such as black holes
\citep[][]{Mashian2017,Breivik2017,Kinugawa2018,Yalinevich2018,Yamaguchi2018,
  Andrews2019}.

Constraining the statistics of opposite ends of the companion mass
function as well as everything in between is crucial to our
understanding of stellar multiplicity which forms one of the
foundations of astrophysics. As a channel to study fragmentation
processes at the birth sites, it informs the theory of star formation
\citep[see e.g.][]{Bate1995,Bate2003,McKee2007}. At high redshifts,
multiplicity of the first stellar systems stipulates how the mass is
apportioned between the Population III stars and thus controls the
ionizing radiation and metal enrichment, which in turn define the
subsequent growth of structure in the Universe
\citep[e.g.][]{Barkana2001,Abel2002,Heger2002,Stacy2010,Stanway2016}. Supernovae
of type Ia are a product of a binary star evolution
\citep[][]{Whelan1973,Tutukov1981,Iben1984,Webbink1984,Maoz2014}, and
several other sub-types are suspected to be
\citep[][]{Podsiadlowski1993,Smartt2009,Smith2011}. Supernovae are not
the only extremely high energy events linked to the binary star
evolution. High mass binaries also serve as progenitors for gamma-ray
bursts \citep[see][]{Narayan1992,Berger2014} and gravitational waves
\citep[see][]{Belczynski2002,Abbott2016}, the two events that in some
cases are also predicted to occur (nearly) simultaneously in the same
system \citep[][]{Blinnikov1984,Abbott2017}. Finally, binaries, even
in very small numbers, control the dynamical evolution of dense stellar
systems \citep[][]{Heggie1975}.

\begin{figure}
  \centering
  \includegraphics[width=0.49\textwidth]{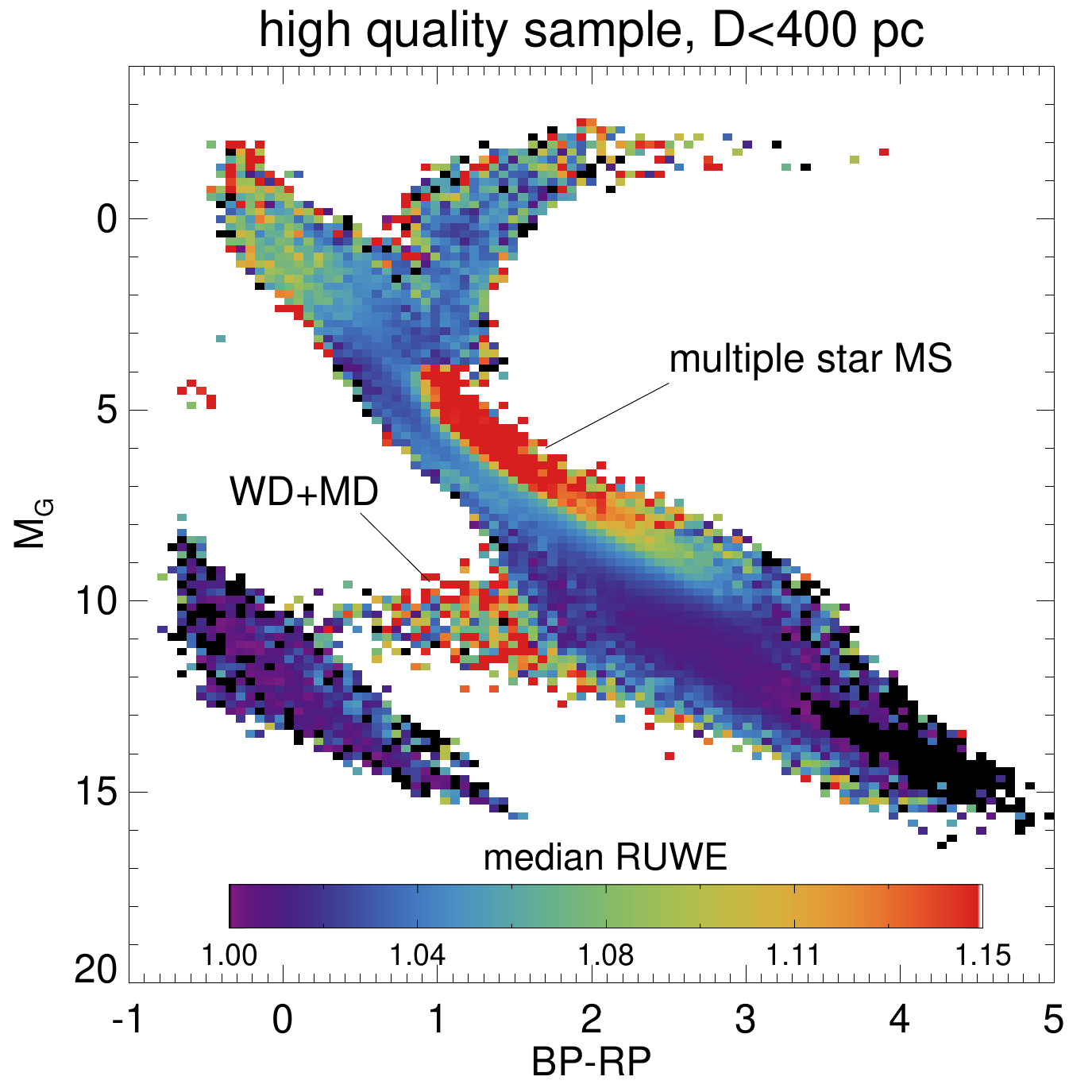}
  \caption[]{Median RUWE as a function of extinction-corrected color
    and absolute magnitude for $\sim4,000,000$ stars selected using
    the criteria listed in Equation~\ref{eq:hrsel} but with a distance
    cut $D<400$ pc. Running parallel to and above the single-star Main
    Sequence is the photometric binary MS which has a notably elevated
    median RUWE. Another region with a clear RUWE excess is the
    sequence of white dwarf-M dwarf binaries at $10<G<12$ and
    BP-RP$<2$. Clear systematic and predictable patterns of RUWE
    variation support the idea of using reduced $\chi^2$ to test the
    presence of unresovled companions to {\it Gaia} stars.}
   \label{fig:ruwe_hrd}
\end{figure}

An impressive variety of observational techniques has been used so far
to probe stellar multiplicity across a wide range of companion masses
and separations. These include photometry, spectroscopy, eclipses,
common proper motions, adaptive optics and interferometry \citep[see
  e.g.][]{Moe2017}. An early example of a comprehensive attempt to
calculate the multiplicity frequency of Solar type stars, including a
correction for observational biases, was reported by \citet{Abt1976}
and updated by \citet{Duq1991}. They used a sample of less than two
hundred stars. Some twenty years later, the analysis was brought up to
date with a sample of about $500$ stars \citep[][]{Raghavan2010} this
time taking advantage of the astrometric distances provided by {\it
  Hipparcos}. These studies not only provided the first robust overall
estimates of the percentages of double, triple and higher-multiple
systems but also detected a clear evolution of the binary fraction
with stellar mass. It is now established that O and B stars are much
more likely to reside in a pair compared to stars further down the
Main Sequence \citep[see][]{Garmany1980,Raghavan2010, Sana2012,
  Duchene2013, Moe2017}. With the advent of wide-angle highly
multiplexed spectroscopic surveys, the sizes of stellar samples
available for the studies of binarity grew by several orders of
magnitude \citep[see e.g.][]{Badenes2012,Hettinger2015,Badenes2018,
  PW2018, EBA2018}. Thanks to the increase in the sample size, trends
in the stellar multiplicity that had previously been hinted at are now
getting firmly established \citep[see
  e.g.][]{PWcirc,Moe2019,PW2020,Merle2020}.

Astrometric surveys in general, and {\it Gaia} in particular, provide
new, complimentary ways of detecting stellar companions. As pointed
out by \citet{Luyten1971}, wide separation binaries can be
straightforwardly identified as pairs of stars with similar distances
and similar proper motions \citep[see][for applications to the {\it
    Gaia} data]{Oh2017,Andrews2017,EBWD2018}. Faint or unresolved
companions can induce a shift of the barycentre with respect to the
photocentre which can be detected when proper motion estimates from
two or more epochs are compared in a method known as the proper motion
anomaly \citep[PMa, see
  e.g.][]{Bessel1844,Brandt2018,Kervella2019,KervellaPuls2019}. Here
we explore a regime complimentary to the proper motion
anomaly. Similar to the PMa method, we study cases where the motions
of the centre of light and the centre of mass are sufficiently
different. If the binary period is smaller than the {\it Gaia}'s's
temporal baseline then the additional centroid perturbation is
non-linear and cannot be absorbed into the proper motion so the
goodness of fit is decreased. This can be detected as an excess in
reduced $\chi^2$.

\section{Photocentre wobble with RUWE}
\label{sec:ruwe}

\begin{figure*}
  \centering
  \includegraphics[width=0.99\textwidth]{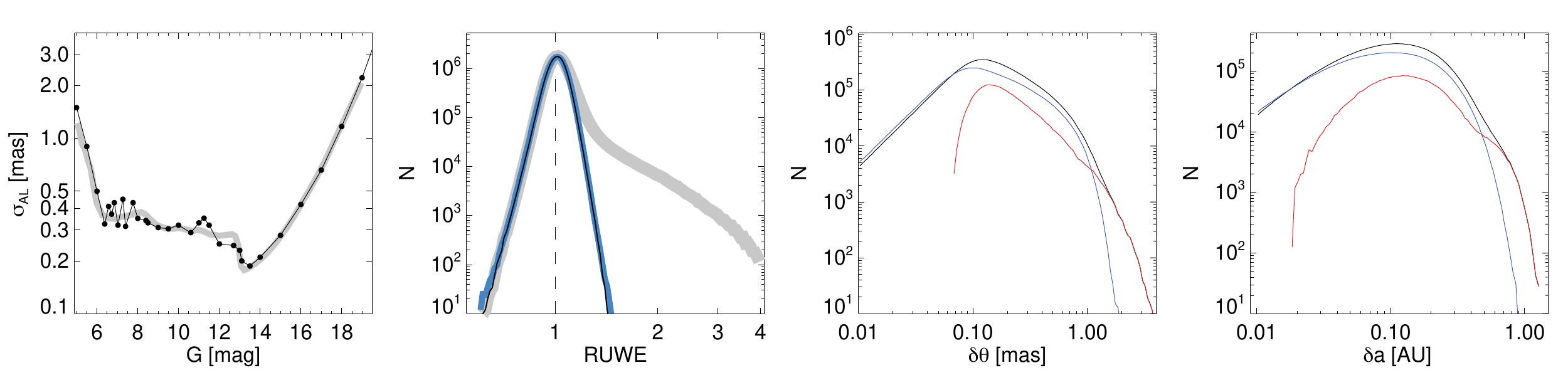}
  \caption[]{Astrometric properties of stars satisfying the selection
    in Equation~\ref{eq:hrsel}. {\it 1st panel:} Black filled circles
    and line show the along-scan error $\sigma_{\rm AL}$ as given by
    the blue line in \citet{Lindegren2018}. Grey curve corresponds to
    $0.53\sqrt N \sigma_{\varpi}$ (see text for details). {\it 2nd
      panel:} Thick grey line gives the Gaia DR2 RUWE distribution for
    the selected sources.  Thin black line corresponds to the RUWE
    distribution reflected around the peak at $\rho_{\rm peak}=1.012$
    which represents the properties of the sources without significant
    centroid perturbation. Blue thick curve is our model of this
    symmetric distribution of unaffected sources (see text for
    details). {\it 3rd panel:} Distribution of $\delta \theta$ for a
    subset of Gaia DR2 sources calculated using
    Equation~\ref{eq:dtheta} (black line) together with our estimate
    of the background (blue line), i.e. the objects with values of
    RUWE close to 1 upscattered by high centroiding errors. Red curve,
    the difference of the two distributions, shows a clear excess of
    sources with noticeable centroid wobble above $\delta\theta\sim0.1$
    mas. {\it 4th panel:} Same as previous panel but for $\delta
    a$. The vertical dashed line in the second panel is the threshold used for the binary
    fraction analysis in Section~\ref{sec:frachrd}}
   \label{fig:error}
\end{figure*}

Our working premise is that the amplitude of the photocentre
perturbation due to binary orbital motion can be gauged from the
reduced $\chi^2$ of the single-source astrometric fit\footnote{Note
  however, that, as we discuss below and demonstrate in detail in
  Penoyre et al (submitted), reduced $\chi^2$ does not always correlate
  with the amplitude of the centroid perturbation, as some of the
  centroid motion can be soaked up in parallax or proper motion
  (creating the so-called proper motion anomaly).}. In practice, we
use a closely related quantity, namely RUWE or $\rho$, the
re-normalised unit weight error \citep[see e.g.][]{Lindegren2018}. The
re-normalisation was required after it was noticed that the peak of
the reduced $\chi^2$ distribution depended on the source colour and
apparent magnitude. Here, we assume that the re-normalisation
\citep[as described in ][]{Lindegren2018} corrects the bulk of the
      {\it Gaia} DR2 systematics so that $\rho^2$ closely approximates
      true reduced $\chi^2$
\begin{equation}
\rho^2\approx\chi_{\nu}^2=\frac{1}{\nu}\sum_{i=1}^{N}\frac{R_i^2}{\sigma_i^2}.
\end{equation}

\noindent Here, $\nu=N-5$ is the number of degrees of freedom, for the
single-source 5-parameter model used in {\it Gaia} DR2. The number of
observations $N=\verb|astrometric_n_good_obs_al|$. $R_i$ and
$\sigma_i$ are the along-scan data-model residuals and the
corresponding centroiding errors of $i$-th measurement of the given
star. If the source is an unresolved binary system, the motions of thef
centre of mass and the centre of light separate. The barycentre motion
is still adequately describable with a 5-parameter model, but the
centre of light trajectory now contains an additional component due to
the binary orbital motion. We therefore expect that unresolved
binaries should yield poorer goodness-of-fit statistics, e.g.\ RUWE.

Figure~\ref{fig:ruwe_hrd} shows the median RUWE value as a function of
the position on the Hertzsprung-Russel Diagram spanned by
extinction-corrected color \bprp\ and absolute magnitude $M_G$ for
$\sim3.87\times10^6$ sources selected using the same criteria as in
Equation~\ref{eq:hrsel} but with a distance cut $D<400$ pc. To remove
the effects of dust reddening we use the maps of \citet{SFD} and
extinction coefficients presented in \citet{Babu2018}. Two sections of
the HRD stand out immediately thanks to a strong RUWE excess indicated
by shades of orange and red. These regions are known to be dominated
by binary stars: the multiple-star Main Sequence that sits above the
single-star MS and the white-dwarf-M-dwarf binary sequence. The clear
pattern of systematic RUWE variation across the HRD as revealed by
Figure~\ref{fig:ruwe_hrd} lends credence to the idea of using the
reduced $\chi^2$ of the astrometric fit to probe for stellar
companions.

\subsection{Amplitude of the angular perturbation $\delta\theta$}

If the photocentre motion deviates from that of a single source, we
can decompose the residual as $R_i=R^\mathrm{ss}_i+\delta\theta_i$, where the
$\delta \theta_i$ represents extra perturbation to the single-source
residual $R^\mathrm{ss}_i$.  We take the root mean square of $\delta
\theta_i$ assuming that the single source portion of $\chi_{\nu}^2$ is
$\sim1$, that $N \gg 5$ and dropping the cross-term, and
interpret the result as the amplitude of the photocentre perturbation (in mas):

\begin{equation}
\delta\theta =\sqrt{<\delta\theta^2_i>}\approx \sigma_{\rm AL}(G)~\sqrt{\rho^2-1},
\label{eq:dtheta}
\end{equation}
Here, we have substituted the per-scan along-scan centroiding error
$\sigma_i$, which is not available, with the mean value $\sigma_{\rm
  AL}$ as a function of source magnitude $G$ presented in
\citet[][blue curve in their Figure~9]{Lindegren2018}.  This is a
robust estimate of the standard deviation of the residuals of the
centroid fit from their residual analysis, not the formal error from
the image parameter determination.  For faint sources, i.e., those
with $G>12$, the difference in the two along-scan centroiding error
estimates is $<20\%$. Note however that for the brighter objects, the
formal error can be some five times smaller than the estimate we chose
to use \citep[see][for details]{Lindegren2018}.  Note that the above
derivation of $\delta\theta$ from $\rho$ is only valid when the binary
motion causes a significant photocentre wobble, i.e., when $\rho>1$.

The first panel of Figure~\ref{fig:error} shows $\sigma_{\rm AL}$ used
here as a function of magnitude $G$. Additionally we demonstrate that
the single-epoch centroiding error can also be estimated as $0.53\sqrt
N \sigma_{\varpi}$, where $\sigma_{\varpi}$ is the reported parallax
error (we have checked that the bulk of the results presented here
does not change if we switch between the two $\sigma_{\rm AL}$
estimates). The second panel of the Figure gives the distribution of
$\rho$ for a sub-set of sources in {\it Gaia} DR2 (grey thick line,
see Equation~\ref{eq:hrsel}). The distribution appears to have two
parts: a peak around $\rho\approx1$ corresponding to single sources or sources
without a measurable centroid perturbation and a tail extending to
large values, corresponding to objects with appreciable centroid
perturbation.

In order to estimate the overall angular photocentre perturbation
corresponding to the tail of this distribution, we construct a simple
model for the RUWE distribution of well-behaved single sources.  We
assume that the distribution of $\rho$ for the unaffected sources is
symmetric, which seems reasonable if the number of observations $N$ is
sufficiently large (the median number of observations for the sample
shown in Figure~\ref{fig:error} is $N=225$) \footnote{Note that even
  for such a high overall number of observations as much as $\sim15\%$
  more sources could be located on the right side of the peak
  (assuming uniform distribution of numbers of observations in the
  range of 150 to 300 with a median of 225).}. We take the $\rho$
histogram and reflect the low-$\rho$ part around the peak $\rho_{\rm
  peak}=1.012$ (thin black line in the second panel of
Figure~\ref{fig:error}).  We approximate this symmetric distribution
as a Student's $t$-distribution with 13.5 degrees of freedom for the
scaled variable $(\rho-\rho_{\rm peak})\delta\rho^{-1}$ where
$\delta\rho=0.057$ is the width of the peak.

\begin{figure}
  \centering
  \includegraphics[width=0.49\textwidth]{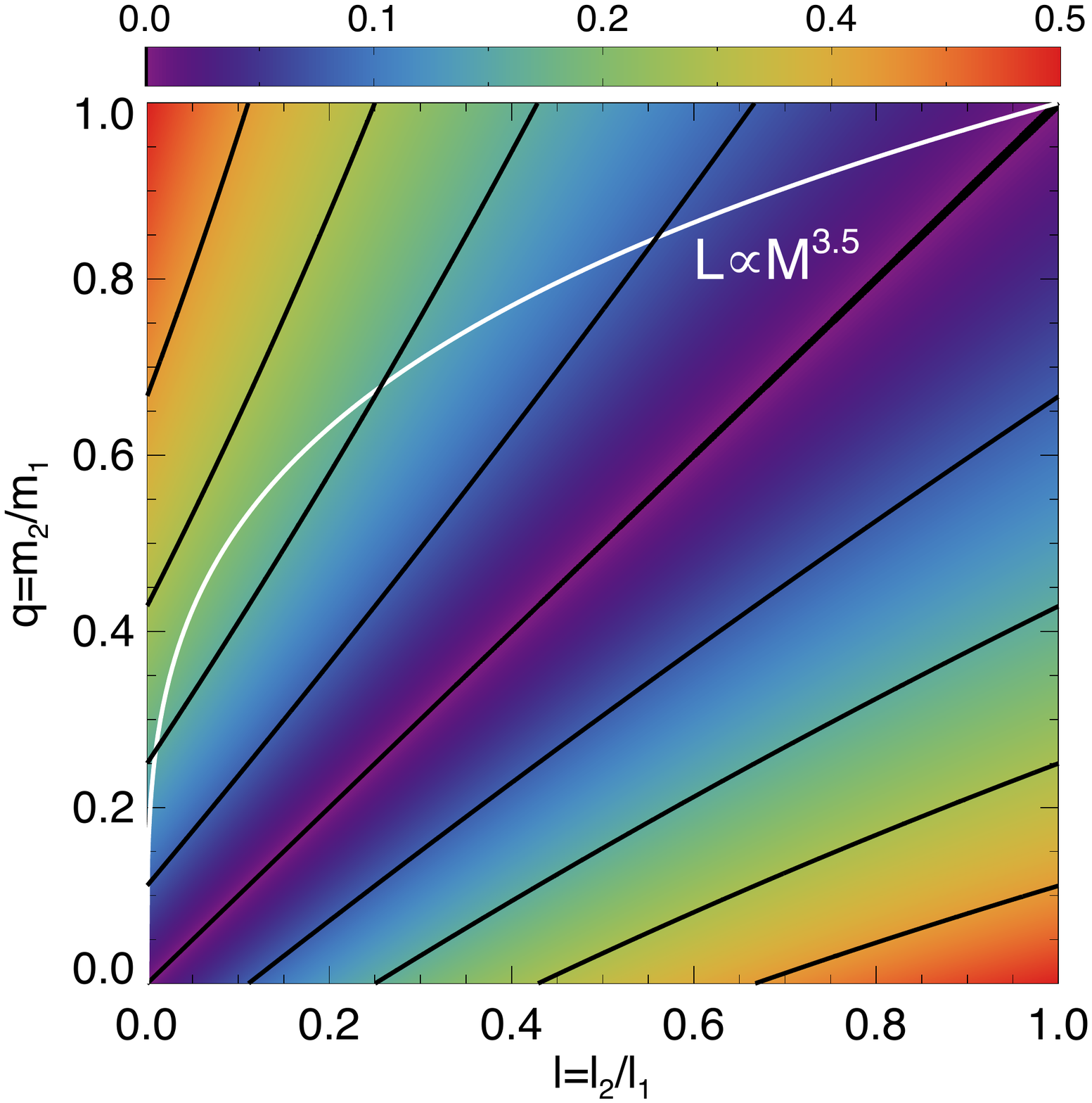}
  \caption[]{Astrometric wobble scaling factor $\delta_{ql}$ as a
    function of the luminosity and the mass ratios, $l$ ($X$-axis) and
    $q$ ($Y$-axis). White curve gives an approximate behaviour for the
    MS stars following a power-law mass-luminosity relation. As
    demonstrated by the MS track, in a stellar binary the typical
    $\delta_{ql}<0.2$}
   \label{fig:delta_simple}
\end{figure}
\begin{figure*}
  \centering
  \includegraphics[width=0.995\textwidth]{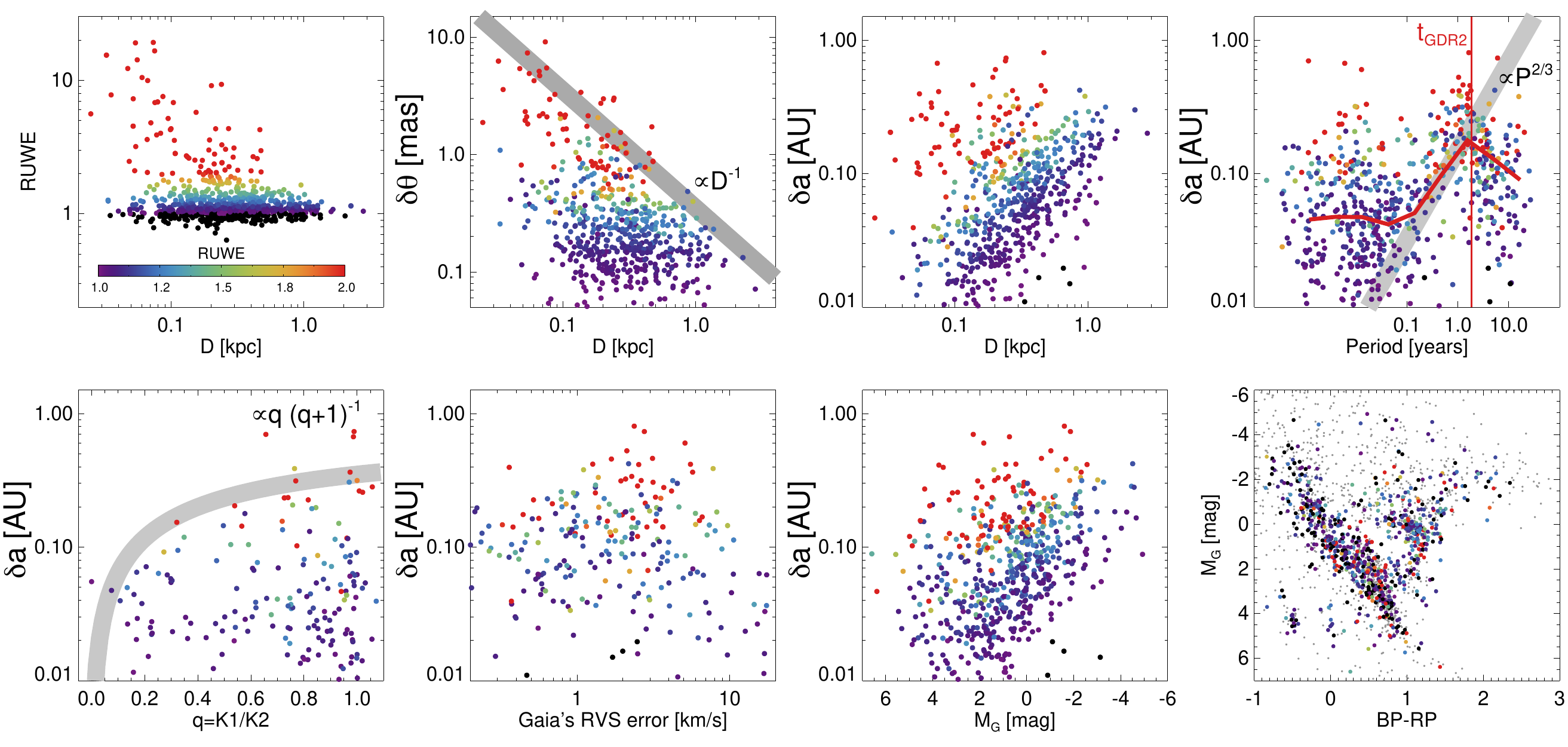}
  \caption[]{SB9 spectroscopic binaries \citep[][]{SB9} as seen by
    {\it Gaia} DR2. Only $801$ binaries satisfying the conditions
    described in the text are shown. {\it Top row, 1st panel:} RUWE as
    a function of distance to the star. The stars are color-coded
    according to the reported RUWE value. This color-coding is
    preserved in all subsequent panels. {\it 2nd panel:} Angular
    displacement in mas $\delta \theta$ as a function of distance. As
    shown by the grey band, the amplitude of the astrometric
    perturbation drops proportionally to the distance. {\it 3rd
      panel:} Physical displacement in AU $\delta a$ as a function of
    distance. Note that the systems of the same separation $a$ induce
    astrometric perturbation of decreasing amplitude with increasing
    distance, thus limiting the {\it Gaia} DR2 sensitivity range to
    2-3 kpc from the Sun. {\it 4th panel:} $\delta a$ as a function of
    the binary period in years. The red curve gives the median $\delta a$
    in a bin of period. Three regimes are apparent. For periods $<1$
    month, the photocentre wobble for distant stars is too low for
    {\it Gaia} to detect robustly. Between 1 month and 22 months, the
    amplitude of the measured photocentre perturbation is proportional
    to the binary's $P^{2/3}$ in accordance with the Kepler's 3rd
    law. Beyond 22 months, {\it Gaia}'s sensitivity drops again as
    this is the DR2's baseline and only a fraction of the induced
    shift is detected by Gaia. Additionally, because of the long-term
    nature of the perturbation, some of the wobble can be absorbed by
    the astrometric solution. {\it Bottom row, 1st panel:} $\delta a$
    as a function of the binary mass ratio $q$. According to
    Equation~\ref{eq:abinary}, the amplitude of the perturbation
    should scale with $q(q+1)^{-1}$ for small $l$. This appears to
    match the behaviour of the upper envelope for those SB9 systems
    with reported K1 and K2, as demonstrated by the grey band. {\it
      2nd panel:} $\delta a$ as a function of the {\it Gaia} RVS
    radial velocity error. Two regimes are discernible: 1) the RV
    perturbation is proportional to the astrometric perturbation and
    2) RV perturbation exceeds astrometric perturbation. This
    demonstrates the complementarity of the two signals. {\it 3rd
      panel:} $\delta a$ as a function of the system's absolute
    magnitude $M_G$. {\it 4th panel:} Hertzsprung-Russel (absolute
    magnitude as a function of color) diagram for the SB9 sources
    colour-coded by their RUWE value.}
   \label{fig:sb9}
\end{figure*}

Using this model for the $\rho$ distribution for single sources we compare
the distribution of angular perturbation $\delta\theta$ of the whole
sample with a control sample composed of single sources only. The
control sample has the size equal to the number of stars in the blue
peak shown in the 2nd panel of Figure~\ref{fig:error}, or, in other
words, twice the number of stars with $\rho<\rho_{\rm peak}$. The
control sample is constructed by pairing random $\rho$ values drawn
from the model single-source $\rho$ distribution described above with
an apparent $G$ magnitude, which gives the corresponding centroiding
error $\sigma_{\rm AL}(G)$.  We calculate $\delta\theta$ using
Equation~\ref{eq:dtheta} for both samples and compare them in the
third panel of Figure~\ref{fig:error}.  The red line shows the
differences between the measured $\delta\theta$ distribution (black)
and the model background (single source) estimate (blue). As the red
line indicates, photocentre wobble with amplitudes as low as $\sim0.1$
mas are detectable.

\subsection{Translation to physical units}

Taking the distance dependence of the centroid wobble into account,
the corresponding physical displacement in AU is:
\begin{equation}
  \frac{\delta a}{\mathrm{AU}}
    = \frac{\delta\theta}{\mathrm{mas}} \frac{D}{\mathrm{kpc}},
\label{eq:da}
\end{equation}
\noindent where $D$ is the distance to the source in kpc computed as
the inverse of parallax in mas. The fourth panel of Figure~\ref{fig:error}
displays the distribution of measured $\delta a$ values as well as our
estimate of the background, i.e., the contribution of sources without
a detectable centroid wobble scattered to high $\delta a$ values (blue
line). The excess of objects with genuine centroid perturbation
(mostly binary stars) is shown with the red line.

We emphasize that our goal is to study the overall binary statistics
with RUWE and the astrometric wobble deduced from it and \emph{not} to
identify individual binary star candidates. It is obvious that at
large distances small individual $\delta\theta$ and $\delta a$ values
(corresponding to $\rho\approx 1$) are not likely to be statistically
significant. Note, however, that closer to the Sun, binary systems
with small separations can yield significant RUWE excess (as discussed
below in Section~\ref{sec:sb9}).  We refrain from identifying binary
star candidates, instead, below, we present evidence for enhanced
binarity for a number of distinct populations of sources. For this, we
rely on two simple methods to gauge the significance of the centroid
perturbation. In some cases (see e.g., Section~\ref{sec:frachrd}), we
use our (admittedly naive) model of single-source $\delta a$ scatter
described above. Elsewhere, we construct comparison samples with
objects whose observed properties (such as apparent magnitude and
color) match those in the population of interest. This allows us to
claim detections of low-amplitude astrometric perturbations when it
shows up as a systematic RUWE excess for the sample as a whole. Here
and elsewhere in the paper we assume that the peak of RUWE
distribution is centred on $\rho\approx1$ by design. This is tested and
shown to be true (in well-populated regions of the CMD) in
Figure~\ref{fig:cmd_ruwe} where we also discuss the behaviour of the
width and the tail of the RUWE distribution. In high-density portions
of the CMD, no strong variations of the RUWE distribution is reported.

An alternative approach could be taken in using \texttt{astrometric
  excess noise} (AEN) as a proxy for $\delta\theta$. It appears
appealing for several reasons: i) AEN was designed precisely to catch
additional perturbation of the stellar photocentre, ii) it does not
include attitude noise and iii) it comes with an estimate of
significance. However, we have decided against using AEN for the
following two reasons. First, AEN ``saturates'' to a zero value for a
large fraction of sources with determined (and reported) RUWE. For
example, for the sample of stars presented in Figure~\ref{fig:error},
only approximately half of sources with RUWE$>1.1$, have AEN$>0$. This
does not pose a problem for sources with large enough perturbations
but limits our understanding of the minimally affected source
(i.e. prevents the calculation of the background model described
above). Second, while the distribution of RUWE is guaranteed (by
design) to peak at 1 across the entire color-magnitude range of {\it
  Gaia}, an equivalent property is not ensured for AEN. It will
therefore contain systematic trends as a function of color and
magnitude, e.g. a strong change around $G\sim13$ due to the d.o.f. bug
in {\it Gaia} DR2 \citep[][]{Lindegren2018}. We have checked the
correspondence between AEN and $\delta \theta$ computed from RUWE and
found them to be strongly correlated (see Figure~\ref{fig:aen}).

\subsection{Astrometric wobble for known binaries}
\label{sec:sb9}

What is the photocentre perturbation expected from binary motion?  In
the limit of unresolved binary with period much shorter than the
observational baseline (such that the orbit sampled over many periods
effectively ends up adding an overall jitter), we can approximate this
as the difference between the center-of-light, which is the
photocentre, and the center-of-mass, which will still follow
single-source astrometric solution.  Given a mass ratio $q=m_2/m_1$
and a luminosity ratio $l=l_2/l_1$, the difference is
\begin{equation}
  \Delta = \left< \left\lvert \frac{\vec{x}_1 + l \vec{x}_2}{1 + l}
    - \frac{\vec{x}_1 + q \vec{x}_2}{1+q} \right\rvert \right>
    = \frac{|q-l| \left< |\vec{x}_1-\vec{x}_2| \right>}{(q+1)(l+1)}
\end{equation}
where the bracket indicates time-average and $\vec{x}_{1,2}$ are the
projected positions of the two stars.  While in detail the relative
vector and how it projects onto the sky plane depends on inclination
and eccentricity (Penoyre et al., submitted), it will be proportional
to the binary separation $a$.  Thus, we expect the following for the
physical size of the wobble $\delta a$:
\begin{equation}
  \delta a\propto\frac{a|q-l|}{(q+1)(l+1)} \equiv a\delta_{ql}
\label{eq:abinary}  
\end{equation}  

\noindent where $\delta_{ql}$ combines the mass and luminosity ratio
factors and determines the link between the actual binary separation
and the measured $\delta a$.  Note that an unresolved binary of two
identical stars ($q=l$) will not show any extra perturbation because
the photocentre coincides with the
center-of-mass. Figure~\ref{fig:delta_simple} shows the behaviour of
$\delta_{ql}$ as a function of the luminosity and mass ratios $l$ and
$q$. White line gives the trajectory for a hypothetical MS population
which follows a power-law mass-luminosity relation. Note that stellar
$\delta_{ql}$ does not exceed $\delta_{ql}=0.5$ because stellar mass
is a monotonic function of stellar luminosity (note however that this
assumption can broken for stars on the RGB and the HB due to mass
loss) and therefore for all luminosity ratios satisfying $0<l<1$, mass
ratios will also remain within $0<q<1$. This however does not hold
true for dim/dark stellar remnants such as white dwarfs, neutron stars
and black holes. For such binary companions, $l\sim0$, while the mass
ratio can be $q\gg1$. If $q$ (or $l$) is allowed to exceed 1, then
$\delta_{ql}$ can exceed 0.5 and reach values close to
$\delta_{ql}\approx1$. We show distributions of $\delta_{ql}$ for
binary, triple and quadruple systems composed of stars drawn from
PARSEC models with different ages and metallicities in
Appendix~\ref{sec:dql}.

\begin{figure*}
  \centering
  \includegraphics[width=0.995\textwidth]{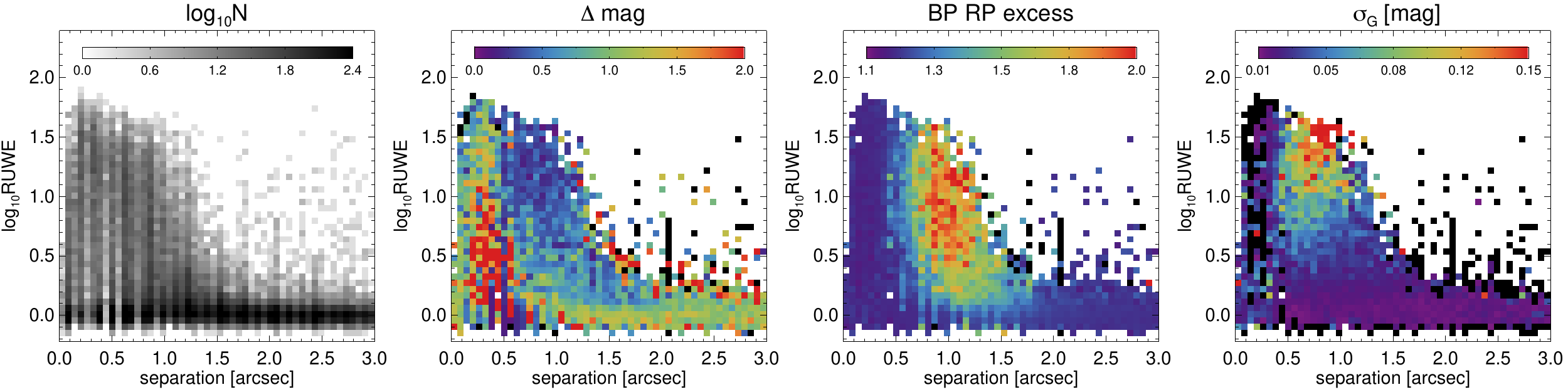}
  \caption[]{{\it Gaia}'s view of the double stars in the WDS
    catalogue \citep[see][]{WDS}. {\it 1st panel:} Logarithm of
    density of sources in the plane of RUWE as a function of binary
    separation (in arcseconds). Many of the WDS systems pile-up around
    $\rho\sim1$. Note however a significant fraction of objects at
    high and extremely high values of RUWE for separations less than
    1.5 arcseconds. While no correlation between RUWE and separation is
    observed, the upper envelopeof the distribution drops with increasing separation.  {\it 2nd panel:} Median magnitude difference of the two
    stars as reported in WDS. Note that even systems with large
    magnitude differences ($\Delta$ mag$>2$) can break {\it Gaia}'s
    astrometry. {\it 3rd panel:} Median \texttt{BP\_RP\_EXCESS\_FACTOR}
    for WDS sources. Comparing with the second panel reveals that {\it
      Gaia} can detect excess flux when the two stars are of
    comparable brightness, i.e. $\Delta$ mag$<0.5$. {\it 4th panel:}
    Median variability amplitude derived from error on the mean flux.}
   \label{fig:wds}
\end{figure*}

Given that the typical $\delta_{ql}$ value is $\sim$0.1 (as
illustrated by the white line in Figure~\ref{fig:delta_simple}),
Figures~\ref{fig:error} and \ref{fig:delta_simple} can be used to
gauge the range of the binary semi-major axes {\it Gaia} DR2 is
sensitive to. The bulk of the $\delta a$ residuals shown in fourth
(right) panel of the Figure~\ref{fig:error} is between 0.01 and 1,
thus implying that most of the detectable binaries will have
$0.1<a$(AU)$<10$. Note that the blue and red curves in
Figure~\ref{fig:error} are given for illustration purposes only
(because they are produced by averaging over all magnitudes and
distances probed). The background $\delta a$ distribution is a strong
function of distance and at low distances its contribution is strongly
reduced, thus allowing for a detection of small separation binaries,
i.e. those with $\delta a\approx0.01$ or even lower. Given the
extended tails of $\delta_{ql}$ and $\delta a$, detection of binary
systems with larger separations, i.e. $a>10$ AU is possible, but for
periods longer than the temporal baseline of DR2, the photocentre
perturbation will become quasi-linear and thus will be absorbed into
the proper motion as illustrated below in the discussion of
Figure~\ref{fig:sb9} (also see Penoyre et al. for detailed
discussion).

\subsubsection{Centroid wobble for SB9 binaries}

Figure~\ref{fig:sb9} tests whether the estimate of the photocentre
wobble derived from RUWE correlates with the properties of
known binary systems. For this comparison, we use the SB9 catalogue
of spectroscopic binaries \citep[][]{SB9}. Presented in the Figure is
the subset of SB9 stars that match to a {\it Gaia} DR2 source within a
1$^{\prime\prime}$ aperture of their reported position. Additionally,
we require
\begin{eqnarray*}
  \verb|PHOT_BP_RP_EXCESS_FACTOR|<3,\\
  E(B-V)<0.5,\\
  \varpi/\sigma_{\varpi}>15,\\
  5<G<20,\\
  D<4~{\rm kpc},\\
  N_{2}=1,\\
  |M_1-G|<1,\\
  {\rm SB9~grade}>1
\end{eqnarray*}  

\noindent here $N_{2}$ is the number of {\it Gaia} DR2 sources within
2$^{\prime\prime}$, $M_1$ is the magnitude of the first component of
the binary as reported in SB9. Only 801 binary systems out of the
total of 2828 recorded in SB9 survive the entirety of the above
cuts. 

\begin{figure*}
  \centering
  \includegraphics[width=0.995\textwidth]{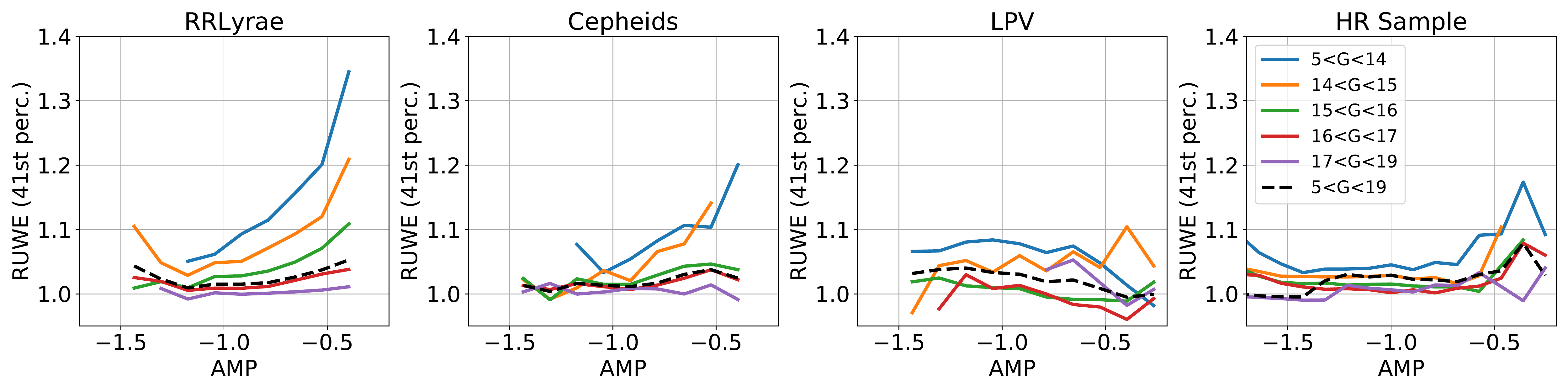}
  \caption[]{RUWE as a function of variability amplitude, estimated
    using Equation 2 in \citet{Belokurov2017}. RR Lyrae (1st panel),
    Cepheids (2nd panel), LPVs (3rd panel) and stars selected using
    criteria specified in Equagtion~\ref{eq:hrsel} are split into 5 groups
    according to their apparent magnitude (colored solid lines, see
    inset in panel 4). For each group, we also show the overall
    behaviour (black dashed line). Bright RR Lyrae and Cepheids show a
    clear correlation between RUWE and variability amplitude.}
   \label{fig:var_ruwe}
\end{figure*}

The first panel of Figure~\ref{fig:sb9} shows the amplitude of RUWE as a
function of distance and demonstrates that the majority of the SB9
sources have RUWE in excess of 1 (in fact, more than $75\%$ of those
that pass the selections cuts listed above do) and that the amplitude
of the RUWE excess increases with decreasing distance. Here, we simply
use $D=\varpi^{-1}$ for the distance estimate. To investigate the
scaling of the astrometric perturbation with distance, it is more
appropriate to convert $\rho$ to $\delta \theta$. Then $\delta \theta$
is expected to decay $\propto D^{-1}$, which is indeed the case as
demonstrated in the second panel of the top row of
Figure~\ref{fig:sb9}. This in turn implies that similar physical
shifts (in AU) would correspond to smaller RUWE excess at larger
distances (see third panel in the top row of the Figure) and thus to
higher contamination. Alternatively, it can be concluded that at small
distances, very tight binary systems can yield significant astrometric
perturbation. For example, in the third panel of the top row in
Figure~\ref{fig:sb9}, there are several objects with $D<0.1$ kpc and
$\rho>2$ (red points) corresponding to $\delta a<0.1$
AU. Extrapolating to lower distances implies that few tens of parsecs
away from the Sun, binaries with separation $\delta a<0.01$ AU can be
studied using {\it Gaia} DR2 astrometry.

The fourth and final panel of the top row of Figure~\ref{fig:sb9}
displays the evolution of the astrometric wobble $\delta a$ as a
function of the binary's orbital period $P$. The red solid line gives
the median $\delta a$ at given period. Three regimes are clearly
discernible here. For periods shorter than 1 month, the astrometric
perturbation can drop below {\it Gaia}'s sensitivity levels
(especially for more distant sources), thus for $P<1$ month, the red
line stays flat around $\delta a \sim 0.05$ AU. For intermediate
values of binary period, i.e. between approximately 1 month and 1
year, the photocentre perturbation, as derived from RUWE, grows
$\propto P^{2/3}$ in accordance with Equation~\ref{eq:abinary} and
Kepler's third law, as indicated by the thick grey band. The median
$\delta a$ curve shown by the red line changes its behaviour abruptly
at $P=22$ months (see vertical thin line). This is the temporal
baseline of {\it Gaia} DR2. For binary systems with periods longer
than $t_{\rm GDR2}$, only a small fraction of the photocentre
excursion is registered by {\it Gaia}. Additionally, at these longer
timescales, some of the centroid shift can be absorbed by the
astrometric solution as an additional (spurious) component of proper
motion (causing the so called proper motion anomaly, see also Penoyre
et al, submitted) and/or parallax. Note that the sharp drop in
sensitivity at 22 months, implies that upper end of the semi-major
axis range probed is dependent on the binary mass.

Switching to the bottom row of Figure~\ref{fig:sb9}, the first panel
gives the dependence of the astrometric wobble amplitude $\delta a$ on
the mass ratio of the binary as measured by the ratio of the velocity
amplitudes of its components. The upper envelope of the distribution
appears to obey Equation~\ref{eq:abinary}, leaving the top left corner
of the Figure empty. The second panel of the bottom row compares the
astrometric perturbation of a binary source to its radial velocity
signature. In practice, we compare $\delta a$ to the radial velocity
error as measured by the RVS on-board {\it Gaia}. The two probes of
binarity are highly complementary, as the RV-based methods are more
sensitive to smaller separation systems where the radial velocity
error (which samples the orbital velocity) reaches higher values,
while the amplitude of the astrometric photocentre perturbation
increases with growing separation. In the second panel of the bottom
row of the Figure, two regimes are visible: at $\sigma_{\rm RV}<5$
km~s$^{-1}$, radial velocity error correlates with $\delta a$. However,
the astrometric signal drops for higher RV perturbations as those, we
hypothesise, are achieved by systems with small separations. The
penultimate panel in the bottom row of Figure~\ref{fig:sb9} shows the
amplitude of the astrometric signal as a function of the source
magnitude. The final (rightmost) panel in the bottom row gives the
positions of the analysed SB9 sources on the HRD. This sample is
dominated by the (young) MS, although many RGB stars are also
present. Interestingly, rather rarer EHBs are also represented (see
the clump at BP-RP$\sim-0.5$ and $M_G\sim$4).

\subsection{Other causes of RUWE excess}

It is not always possible to relate the quality of the {\it Gaia}'s
astrometric fit to the physical properties of the binary. In what
follows we consider two such cases.

\subsubsection{Marginally resolved sources}
\label{sec:wds}

\begin{figure*}
  \centering
  \includegraphics[width=0.995\textwidth]{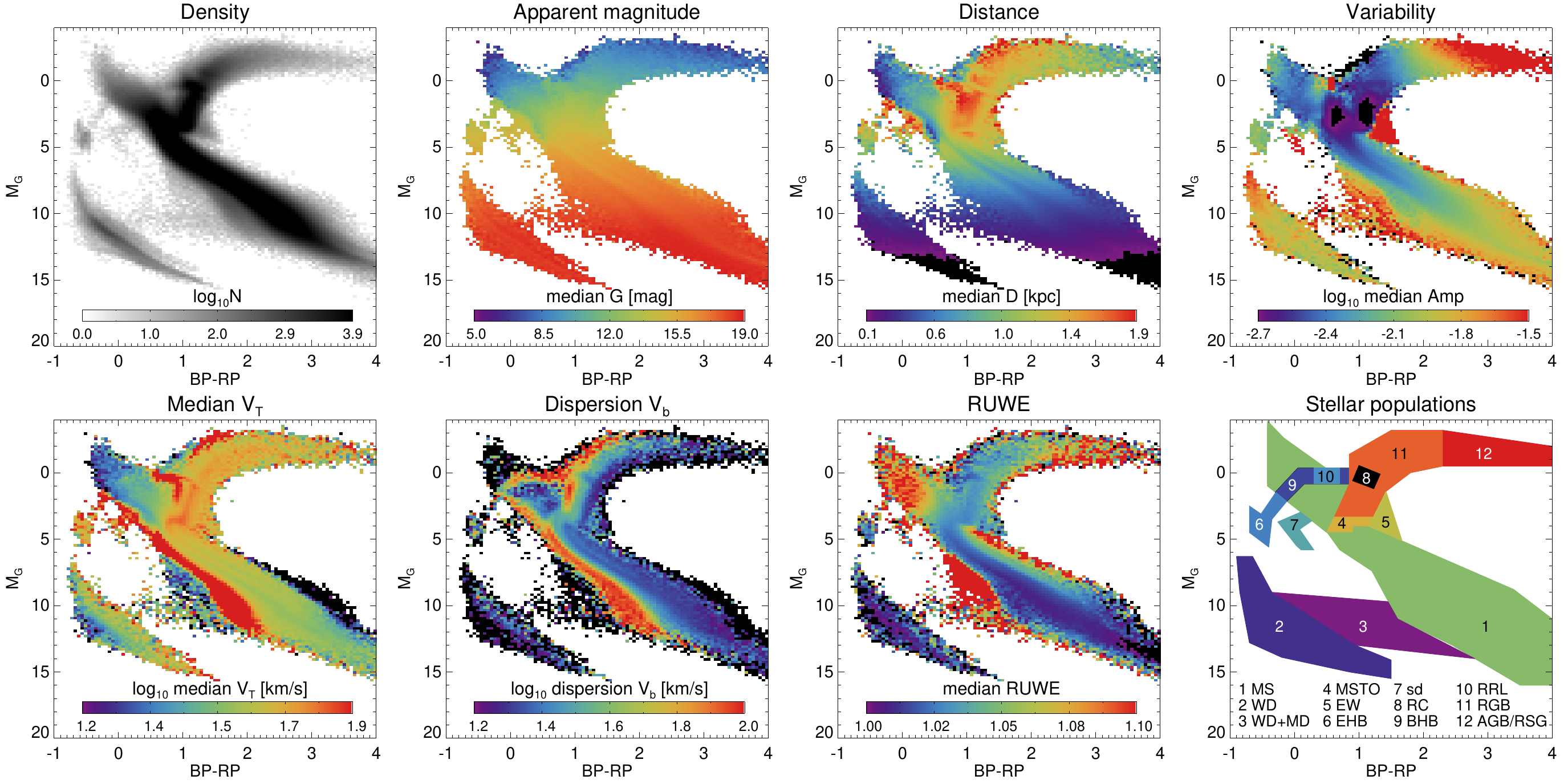}
  \caption[]{Hertzsprung-Russel Diagram for $\sim2.2\times10^7$ {\it
      Gaia} DR2 sources satisfying the selection criteria described in
    Section~\ref{sec:frachrd}. {\it Top row, 1st panel:} Logarithm of
    source density, pixel size is 0.053$\times$ 0.195 mag. {\it 2nd
      panel:} Median extinction-corrected apparent magnitude. Note a
    strong correlation as a function of absolute magnitude $M_G$. {\it
      3rd panel:} Median heliocentric distance. A pronounced gradient
    as a function of both colour and magnitude is visible. {\it 4th
      panel:} Median variability amplitude (see
    Section~\ref{sec:frachrd} for details). A complicated patchwork of
    regions with significant variability is noticeable. {\it Bottom
      row, 1st panel:} Median heliocentric velocity $V_T$. The HRD
    space can be seen separated into thin disc (blue) thick disc
    (yellow) and halo (red) populations. {\it 2nd panel:} Similar but
    entirely the same pattern can be seen when the HRD is colored by
    the dispersion in latitudinal proper motion corrected for the
    Solar reflex. {\it 3rd panel:} Median RUWE. Several regions with
    elevated $\rho$ are apparent. These include the photometric binary
    MS, B stars, and the reddest portion of the AGB. Additionally, a
    portion of the HRD below the MSTO exhibits elevated levels of
    RUWE. We argue that the astrometric solutions for these stars are
    completely broken. {\it 4th panel:} Approximate stellar population
    boundaries.}
   \label{fig:hr}
\end{figure*}

The most obvious such situation occurs when the double star is nearly
resolved, i.e. the stellar image is perturbed from a single-star PSF
but {\it Gaia} identifies and measures it as a single object. As a
result, there exists a gross mismatch between the image shape and the
PSF/LSF model of it, which results in a large centroiding error. The
nominal ``centre'' of such semi-resolved binary image depends strongly
on the scan angle and will oscillate wildly as a function of
time\footnote{That partially resolved sources can lead to unreliable
  astrometry is also mentioned in the considerations on the use of
  Gaia DR2 astrometry
  (\url{https://www.cosmos.esa.int/web/gaia/dr2-known-issues}). See
  slide 48 of the associated presentation by Lindegren et al.}. We
explore the details of this catastrophic break-down in
Figure~\ref{fig:wds} which shows {\it Gaia}'s astrometry for sources
in the Washington Double Star (WDS) catalogue \citep[][]{WDS}. The
first panel of the Figure shows the logarithm of the density of stars
in the plane spanned by RUWE and the separation of the double star. A
sharp climb-up of RUWE to extreme values is observed for separations
less than $\sim1.5$ arcsec. However, as obvious from the Figure there
is no correlation between $\rho$ and the binary separation. As the
second panel demonstrates, large RUWE values are reported for a wide
range of the companion magnitude difference: at small separations,
even faint companions can cause significant centroid displacement. The
brighter companions can possibly be picked up as they contribute to a
noticeable BP/RP excess as illustrated in the third panel of the
Figure. Finally, the fourth panel shows that at separations $>0.5$
arcsec, semi-resolved objects start to show significant variability
(as gleaned by the error of mean flux measurement). Overall, the bulk
of semi-resolved double-stars can in principle be filtered out by
applying cuts on BP/RP excess and variability. Note however, that
double-stars with separations less than $\sim0.5$ arcseconds can not
be identified this way. As the second panel of the Figure illustrates,
at these separations, RUWE appears to grow proportionally to the
luminosity ratio. It is therefore possible that in this regime, RUWE
scales similarly to that for unresolved binaries. In what follows, we
do not attempt to cull potential semi-resolved double-stars (although
a cut on parallax error gets rid of most of extreme RUWE cases). This
should not affect the analysis presented below under the assumption
that semi-resolved double-stars affect all stellar populations
equally.

\subsubsection{Variability}
\label{sec:var}

We also detect a tendency for RUWE to increase slightly with
photometric variability as illustrated in Figure~\ref{fig:var_ruwe}
for RR Lyrae, Cepheids \citep[see][]{Clementini2019} and Long Period
Variables \citep[see][]{Mowlavi2018}, as well as the sample of objects
satisfying the cuts presented in Equation~\ref{eq:hrsel}. The RR Lyrae
stars have been selected joining the objects classified as RR Lyrae in
the {\it Gaia} SOS \citep[Specific Object Studies,
  see][]{Clementini2019} and general variability \citep[see][]{Holl18}
tables. Stars in known globular clusters, dwarf satellites, Magellanic
Clouds and Sagittarius stream have been removed following
\cite{IorioBelokurov19}. Finally, we apply quality cuts in all the
three variable star catalogs filtering stars on $|b|$,
$\verb|PHOT_BP_RP_EXCESS_FACTOR|$ and $E(B-V)$ as in
Equation~\ref{eq:hrsel}. The final cleaned catalogs contain 4163
Cepheids, 41317 RR Lyrae and 10016 Mira stars.  The variability
amplitude is estimated using equation 2 of \citet{Belokurov2017} and
we checked that it nicely correlates with the peak-to-peak light curve
amplitude measured by {\it Gaia} for a subsample of stars in our
selection.

For each group of stars, individual panels of
Figure~\ref{fig:var_ruwe} show the evolution of RUWE as a function of
the photometric variability amplitude in 5 apparent magnitude bins
(colored solid curves) as well as for the whole group (black dashed
curve).  The strongest signal is exhibited by the bright RR Lyrae
$5<G<14$: for these, RUWE correlates strongly with the variability
amplitude and can reach $\rho\sim1.3$ at the high end. This
correlation subsides for fainter RR Lyrae, while the faintest stars
with $17<G<19$ do not show any connection between RUWE and
variability. Similarly, RUWE values for the bright Cepheids,
i.e. those with $5<G<14$ and $14<G<15$ appear to correlate with
amplitude, but this dependence disappears for stars fainter than
$G=15$. Note that for (all) bright sources RUWE could be affected by
the systematics in centroiding and attitude calibrations. The LPV RUWE
values show no correlation with amplitude (3rd panel), with a very
slight overall increase for stars brighter than $G<15$. The
high-quality sample (Equation~\ref{eq:hrsel}, 4th panel) shows the
lowest overall values of RUWE at all $G$, i.e. $\rho\sim1$, apart from
the highest amplitude sources. The cause of the dependence of RUWE on
variability as a function of magnitude for some objects may be
understood by inspecting Figure 6 in \citet{RUWE}. The normalization
coefficient $u_0$ is a strong function of both apparent magnitude and
color, experiencing sharp changes at a several values of $G$,
e.g. $G\sim13$. A variable object will be measured by {\it Gaia} at a
range of magnitudes (and colors) and therefore its RUWE can not be
normalized using a single value of $u_o$. This spurious induced RUWE
excess will be worse for the stars whose variability takes them across
the sharp features seen in Figure 6 of \citet{RUWE}. An additional
contribution to the RUWE for variable stars is due to the assumption
that the PSF used in the centroiding of the source images is
independent of source colour and magnitude \citep[section 2.2
  in][]{Lindegren2018}. These effects make it difficult to interpret
any RUWE excess for variable stars.

\begin{figure*}
  \centering
  \includegraphics[width=0.995\textwidth]{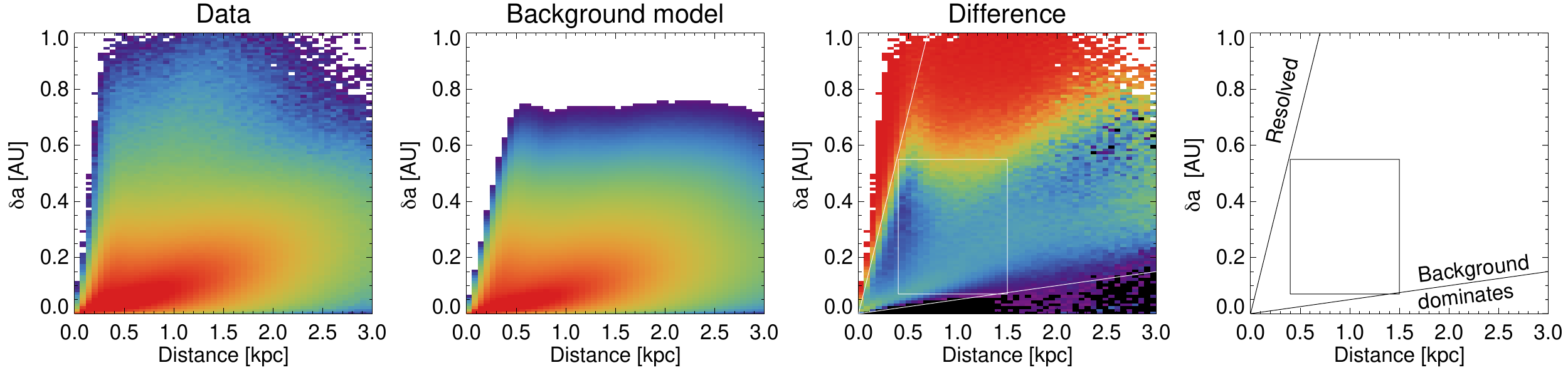}
  \caption[]{{\it 1st panel:} Logarithm of source density in the space
    of centroid wobble $\delta a$ and distance for a subset of Gaia
    DR2 objects selected using equation~(\ref{eq:hrsel}). {\it 2nd
      panel:} Same as previous panel but for a mock background
    sample. {\it 3rd panel:} Difference between the measured and
    background density distribution of $\delta a$ as a function of
    distance. Rectangular box shows the selection boundaries used to
    study binary fraction in Section~\ref{sec:frachrd}. Diagonal lines
    mark the regimes where i) the binary sources stop being resolved
    by Gaia and ii) the background starts to dominate as the scatter
    is amplified by a factor $\propto D$. {\it 4th panel:} Selection
    boxes (see previous panel) only shown for clarity.}
   \label{fig:compl}
\end{figure*}

\section{Some applications}

\subsection{Binary fraction across the Hertzspung-Russell Diagram}
\label{sec:frachrd}

We explore how the binary fraction evolves across the HR diagram as
traced by RUWE. Figure~\ref{fig:hr} presents the distribution of
$2.2\times10^7$ {\it Gaia} sources in the space of
extinction-corrected absolute magnitude $M_G$ and colour \bprp. These
objects were selected by applying the following criteria.

\begin{equation}
  \begin{aligned}
  |b|>15^{\circ},\\
  \verb|PHOT_BP_RP_EXCESS_FACTOR|<3,\\
  E(B-V)<0.25,\\
  \varpi/\sigma_{\varpi}>10,\\
  \sigma_{\varpi}<\sigma^{97}_{\varpi},\\
  5<G<19,\\
  0.01<D<3~{\rm kpc},\\
  N_{8}=1.\\
  \end{aligned}
  \label{eq:hrsel}
\end{equation}

\noindent Here $N_{8}$ is the number of {\it Gaia} sources within an
8$^{\prime\prime}$ aperture and $\sigma^{97}_{\varpi}$ is the 97th
percentile of the parallax error distribution in a given (uncorrected
for extinction) apparent magnitude $G$ bin. The motivation for the
$\sigma^{97}_{\varpi}$ cut is to cull objects with the worst
astrometric solutions, especially those whose parallaxes we can not
trust (we suspect that some of these could be semi-resolved blended
stars discussed in Section~\ref{sec:wds}).

The first panel in the top row shows the logarithm of the HRD density
distribution. Many familiar ubiquitous, as well as rare, stellar
populations are readily identifiable. These are (going from faint to
bright stars) the main sequence (MS, 1), the white dwarf (WD, 2)
sequence, the White-dwarf-M-dwarf binary sequence (WD+MD, 3), the MS
turn-off (MSTO, 4), the contact eclipsing binaries (EW, 5), the
extreme horizontal branch (EHB, 6), sub-dwarfs (sd, 7) , the red clump
(RC, 8), the blue horizontal branch (BHB, 9), the RR Lyrae (RRL, 10),
the red giant branch (RGB, 11) and the asymptotic giant branch (AGB,
12). As the second panel in the top row of Figure~\ref{fig:hr}
illustrates, there is a strong correlation between the position on the
HR diagram and the apparent magnitude of a star. Also, given a wide
range of intrinsic luminosities, stars in different portions of the HR
space, reach different distances from the Sun (third panel of the top
row). Given that the observed astrometric perturbation is proportional
to the apparent magnitude (via $\sigma_{\rm AL}$) and inversely
proportional to distance, these strong couplings may imprint
significant selection biases in the distribution of RUWE across the HR
space.

\begin{figure*}
  \centering
  \includegraphics[width=0.995\textwidth]{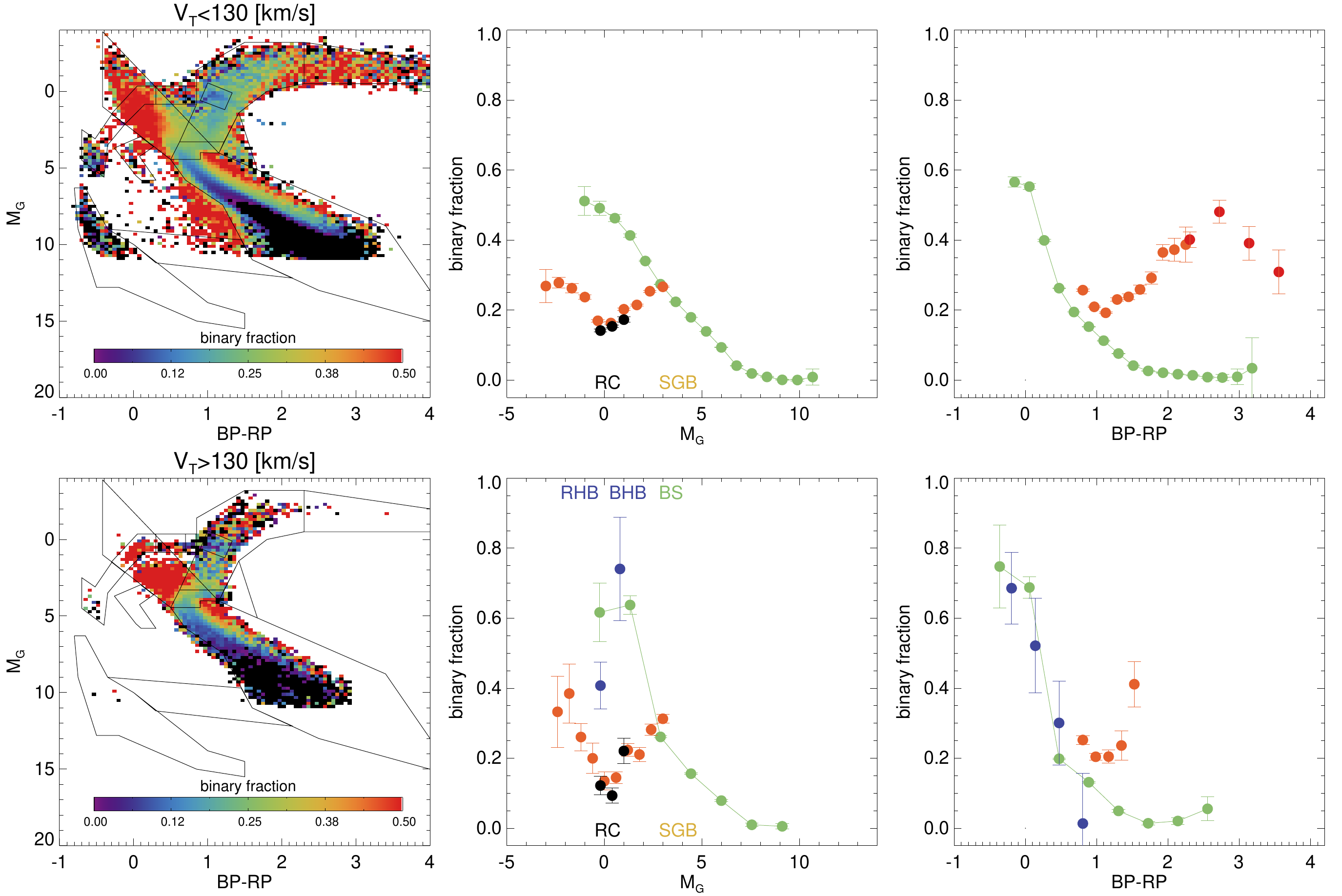}
  \caption[]{Binary fraction across the HRD. {\it Top:} Low
    heliocentric velocity sample $V_T<130$ kms$^{-1}$. {\it Bottom
      row:} Same as the top row but for stars with high heliocentric
    tangential velocity $V_T>130$ kms$^{-1}$. {\it Left column:} Ratio
    of the number of stars within the selection box shown in panels 3
    and 4 of Figure~\ref{fig:compl} to the total number of stars in
    the pixel. Black lines show the same stellar population boundaries
    as shown in Figure~\ref{fig:hr}. Several regions in the HRD show a
    clear excess of binary stars: i) the photometric binary (and
    higher multiples) sequence, ii) the young MS. On the other hand,
    photometric single-star MS shows low binary fraction. Binarity is
    also subdued on the RGB. Finally, for the high tangential velocity
    sample (shown in the bottom row), BS and BHB regions show a clear
    and strong binary fraction enhancement. {\it Middle column:}
    Background-subtracted binary fraction as a function of absolute
    magnitude $M_G$.  Note i) a trend of increasing binary fraction
    with increasing stellar luminosity on the MS, ii) significantly
    lower fraction on the RGB and iii) a dip in binarity around the
    RC. {\it Right column:} Same as previous column but as a function
    of colour BP-RP.}
   \label{fig:hr_da}
\end{figure*}

Many additional correlations are apparent. For example, shown in the
fourth (and final) panel of the top row of Figure~\ref{fig:hr} is the
median variability amplitude as gauged by the mean flux error
\citep[see][]{Belokurov2017}. Note that for this plot we convert the
amplitude into magnitudes and subtract the median magnitude error in
quadrature as a function of $G$. Here, three regions dominated by high
amplitude of variability are apparent: contact eclipsing binaries
(EW), RR Lyrae and long-period variables (LPV) and Mira stars. As
illustrated in the first and second panels of the bottom row of the
Figure, stars also cluster differently in the HR diagram depending on
their kinematics. High heliocentric tangential velocity or large
dispersion in the latitudinal component of the proper motion tends to
pick up the thick disc and halo populations. These stars are typically
older and more metal-poor than the thin disc and thus pile-up on the
blue side of the MS and the RGB. Also, high tangential motion
selection tends to emphasise the horizontal branch stars, typical
denizens of the halo (and possibly thick disc). The third panel of the
bottom row of Figure~\ref{fig:hr} shows the median RUWE in pixels of
the HRD. Apart from the stars with spurious parallax measurements
directly underneath the MS at $1<\bprp<1.5$, there are three regions
of the HRD where median RUWE is significantly different from $\rho=1$:
the AGB, the YMS and the binary (and ternary) MS.  At faint
magnitudes, the WD sequence stands out as the region of the HRD with
the lowest fraction of stars with RUWE excess. This should not be
surprising: both of the progenitors in the WD binary must have evolved
away from the MS and expanded while ascending the RGB. In close
systems this would result in interaction and merging. Wide double
white dwarfs would not have interacted but these have periods longer
that what {\it Gaia} DR2 is sensitive to ($\sim2$ years or
larger). Between the WD+MD sequence and the MS there exists a region
with very high RUWE values. We have investigated the properties of
these stars and concluded that their parallax measurements likely
suffer a strong systematic bias. Based on their proper motions, they
are likely to be distant MS stars for which {\it Gaia} overestimates
parallax significantly. This could happen because some of these stars
are partially resolved double stars (see
Section~\ref{sec:wds}). Additionally, some of these could be binary
systems with the orbital periods close to 1 year (see the companion
paper by Penoyre et al, submitted). Finally, the fourth (rightmost)
panel of the bottom row presents a combination of masks marking the
locations of the stellar populations mentioned above.

\subsubsection{A binary fraction estimate}

We estimate the fraction of stars in binary systems by calculating the
number of objects with a centroid perturbation above a certain
threshold. This ought to be done with $\delta a$ because RUWE and
$\delta\theta$ strongly depend on the apparent magnitude and
distance. Figure~\ref{fig:compl} helps to understand the sensitivity
range of {\it Gaia} DR2. It shows the source density in the plane of
centroid perturbation (in AU) $\delta a$ as a function of distance (in
kpc). The observed distribution can be compared to the model
distribution of single (or unperturbed) stars shown in the second
panel of the Figure. The third panel displays the density difference
between the data and the background model. Two regimes are immediately
apparent.  At small distances, large separation binaries are resolved
and therefore an empty triangular-shaped region can be seen in the
left portion of all three panels. At small separations, it is
progressively difficult to detect sources with statistically
significant centroid perturbations at larger distances. Hence, a dark
region (with negative residuals) can be seen in the bottom right of
the third panel. Guided by these trends, we select a region with
$0.4<D ({\rm kpc})<1.5$ and $0.07<\delta a ({\rm AU})<0.55$, which we
use to estimate the binary fraction. This particular range of $\delta
a$ is also beneficial as it allows us to probe the regime where the
bulk of the sources lie (for the given distance bracket), compared to
say $\delta a>0.5$ where the sensitivity is nominally higher but where
very few objects exist.

Figure~\ref{fig:hr_da} shows the fraction of stars falling within the
selection boundaries marked in the third and the fourth panels of
Figure~\ref{fig:compl} for each pixel of the HRD. Note that the number
of stars satisfying the $\delta a$ and heliocentric distance cuts
stated above is corrected for the background contribution, which
requires an estimate of the number of unperturbed sources. The number
of unperturbed sources is calculated for each pixel of HRD as twice
the number of stars with $\rho<\rho_{\rm peak}$, where $\rho_{\rm
  peak}$ is the same as above. For the low tangential velocity, and
therefore more metal-rich and younger, stellar population shown in the
top row of Figure~\ref{fig:hr_da} three distinct areas of elevated
binary fraction are visible. They are the photometric binary main
sequence running parallel to and above the single star MS, the young
MS at $\bprp<0.5$ and the AGB region. For the high tangential
velocity, and therefore more metal-poor and older, stellar population
shown in the lower row of the Figure, the binary faction similarly
increases at the binary MS, which is offset to the blue compared to
its metal-rich counterpart shown in the top row. Both Blue Horizontal
Branch stars and Blue Stragglers show highly elevated binary
fractions. Along the RGB, for both slow and fast $V_T$ stars the
binary fraction remains approximately constant. In both rows, a clear
drop in binarity is noticeable around the RC location. For redder
giant stars, i.e. those with $BP-RP>2$ which end up in our AGB box,
the binary fraction shows a mild increase. Perhaps, the simplest
explanation of this signal is that the RUWE excess is spurious and is
caused by stellar variability (as discussed in
Section~\ref{sec:var}). Indeed many of the stars in this part of the
HRD and Long Periodic Variables.

The middle (right) column of the Figure presents binary fraction as a
function of absolute magnitude (colour) for the MS (RGB) in green
(orange). Several trends are immediately visible. First, along the MS
the binary fraction grows as a function of decreasing $M_G$ and
$BP-RP$, indicating an increase in binarity with stellar mass. On the
RGB, the binary fraction is approximately half that of the MS at
similar luminosities. A region around the RC location shows a
noticeable dip in binarity compared to the rest of the RGB.

Superficially, the trends in our binary fraction estimates match those
reported in the literature. We see a continuous evolution where the
binary fraction peaks for the high-mass MS stars, declines for the
Solar-like stars and drops quickly for the MS dwarfs
\citep[c.f.][]{WardDuong2015}. Let us re-iterate however that the
reported binary fractions should only be assessed in relative terms.
These fractions are always lower (by an amount which is only known
approximately) compared to the published estimates due to the
selection bias introduced by design: we measure the incidence of
systems with separations above a certain threshold.

\begin{figure*}
  \centering
  \includegraphics[width=0.99\textwidth]{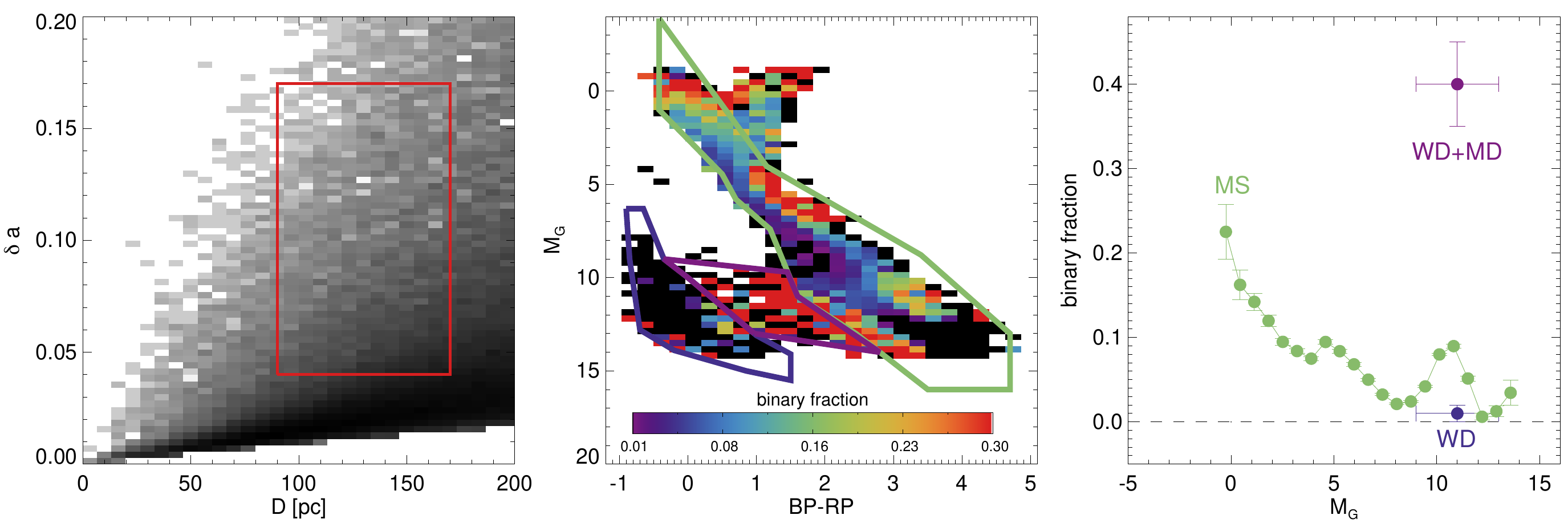}
  \caption[]{{\it Left:} Photocentre perturbation $\delta a/{\rm AU}$
    as a function of distance (in pc) for the same sample considered
    in Figures~\ref{fig:compl} and~\ref{fig:hr_da} but closer to the
    Sun. Red rectangle gives the selection boundary used to compute
    the binary fraction shown in the next two panels. {\it Middle:}
    Fraction of stars with $0.04<\delta a/{\rm AU}<0.17$ amongst those
    with $90<D/{\rm pc}<170$ as a function of position on the
    Hertzsprung--Russell diagram. {\it Right:} Overall binary fraction
    for the MS stars as a function of absolute magnitude $M_G$
    (green). Also shown the binary fraction measurements for the
    double WDs and WD+MD pairs.}
   \label{fig:fdwarf}
\end{figure*}

\subsection{White and Red Dwarfs}

The selection discussed above is limited to distances in excess of
$D=0.4$ kpc and therefore excludes intrinsically faint stars such as
white dwarfs and M dwarfs. Here we analyze a smaller volume limited to
$90<D/{\rm pc}<170$ which allows us to estimate the incidence of WD,
MD and WD+MD pairs. At these distances, wide pairs are resolved, but
conversely, {\it Gaia}'s astrometry is sensitive to binaries with
smaller orbits, therefore we count objects with separations
$0.04<\delta a/{\rm AU}<0.17$.

The left panel of Figure~\ref{fig:fdwarf} shows the
background-subtracted distribution of $\delta a$ as a function of
heliocentric distance. For the stars abiding by the selection criteria
specified in Equation~\ref{eq:hrsel} and lying in the distance range
discussed above (see the boundary shown in red in the left panel) the
middle panel of the Figure gives the map of the binary fraction as a
function of colour and absolute magnitude. The third panel of the
Figure presents binary fraction estimates for the three stellar
populations highlighted in the middle panel, the MS, the WD and the
WD+MD pairs. In agreement with the measurement for the more distant
sample presented in the previous section the MS binary fraction
increases steadily with $M_G$ (and hence with mass). Using the local
sample, we can extend the trend to $M_G>10$. This reveals two wiggles
in the binary fraction curve, one at $M_G\sim 5$ and another one at
$M_G\sim 10$. Some of the increase at $M_G\sim10$ is likely due to
WD+MD pairs where the white dwarf companion is cool and faint enough
not to change the colour of the MS companion drastically. It is
unclear if the entirety of binary fraction fluctuation can be
explained by the WD+MD pairs. As a population, the photometric WD+MD
pairs residing in the area of the HRD between the MS and the WD
sequence posses the highest binary fraction in the sample
considered. Note that, obviously, nearly all objects\footnote{There
  could be some contamination from objects with spuriously large
  parallaxes as discussed above.} in this region of colour and
magnitude should be binaries. Note that as above, our estimate
concerns the systems in a particular range of semi-major axis sizes.

The incidence of WD+WD binaries is the lowest at $\sim1\%$. This is in
good agreement with the recent measurement of \citet{Toonen2017} who
report the binary fraction of $1\%-4\%$ for their unresolved double
WDs. Assuming that all objects within the WD+MD mask are binaries, the
observed fraction of $\sim40\%$ can be used to estimate the selection
bias for double WD in our sample, which gives $2.5\%$ after
correction.

\subsection{Blue Stragglers}

The left panel in the bottom row of Figure~\ref{fig:hr_da} reveals a
substantial population of stars between the turn-off and the
horizontal branch that are likely too luminous and too hot for the
typical age of the selected sample. These are the so-called Blue
Lurkers or Blue Stragglers
\citep[][]{Sandage1953,Burbidge1958}. According to current theories,
more than one star is needed to make a Blue Straggler. In dense
stellar systems, such as globular clusters, the suspiciously
young-looking BSs are probably made by direct stellar collisions
\citep[][]{Hills1976}. Note that the frequency and the efficiency of
such interactions can be greatly enhanced if the colliding systems are
binaries to begin with \citep[][]{Leonard1989,Bailyn1995}. Thus, even
in star clusters, the bulk of BSs have probably come from a parent
binary population
\citep[see][]{Leigh2007,Knigge2009,Geller2011}. Alternatively,
irrespective of its environment, a star in a binary system can be
rejuvenated as a result of the mass transfer from its companion
\citep[][]{Mccrea1964, Chen2008a,Chen2008b}. Finally, the third
scenario invokes a parent triple system in which the Kozai-Lidov
mechanism pushes the inner binary to undergo Roche-lobe overflow and
possibly merge \citep[][]{Perets2009}. It is likely that a combination
of the above mechanisms is required to explain the observed properties
of BSs. Note that in all three scenarios, the BSs can be either a
stellar merger product or a result of a mass transfer. The latter
yields a BS in a binary, while the former can be a single star.

\begin{figure}
  \centering
  \includegraphics[width=0.495\textwidth]{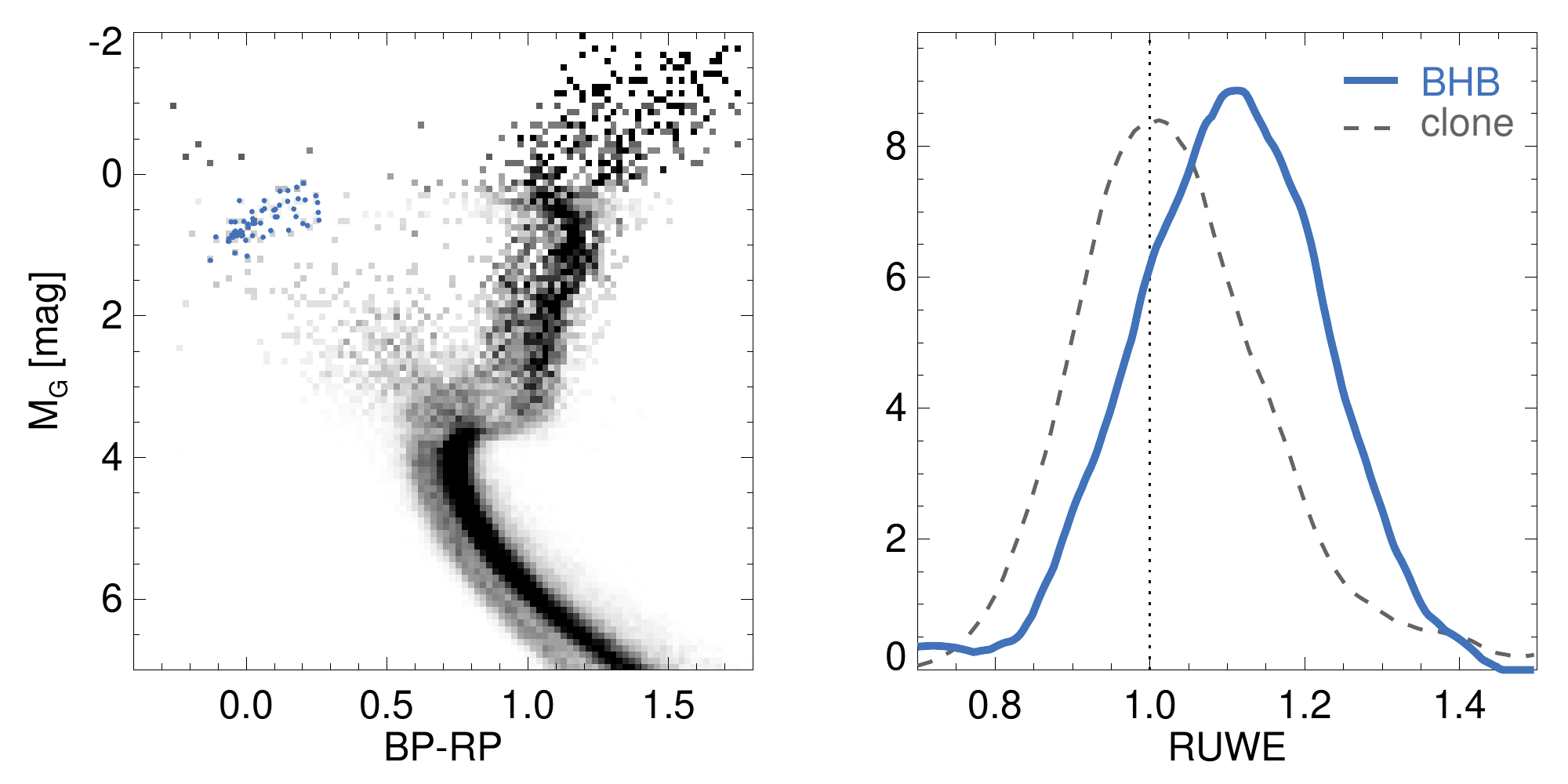}
  \caption[]{Photocentre wobble of Blue Horizontal Branch stars. {\it
      Left:} Color-Magnitude Diagram of high heliocentric tangential
    velocity stars (see Figure~\ref{fig:hr_da}). BHBs stand away from
    the rest of the old stellar populations due to their relatively
    high temperatures and intrinsic luminosities. Selected BHB
    candidates are shown in blue. {\it Right:} Distribution of the
    RUWE values for the BHB candidates selected as shown in the left
    panel (solid blue) and the comparison sample (`clone'; dashed
    grey). The `clone' sample contains stars of marching \bprp\
    colour and apparent magnitude $G$ (before extinction correction).}
   \label{fig:bhb}
\end{figure}

According to Figure~\ref{fig:hr_da}, the binary fraction in the BS
region of the HRD is the highest across the entire sample. Given the
possible contamination and the systematic biases associated with our
simple measurement procedure, it is quite likely that we underestimate
the true binary fraction amongst the BS. These could be 100\%
binaries. Generally, our measurements are in agreement with the
earlier spectroscopic studies that found a high fraction of binaries
amongst the field BSs \citep[e.g.][]{Preston2000,
  Carney2001,Carney2005,Jofre2016,Matsuno2018} and BSs in star
clusters \citep[e.g.][]{Mathieu2009}. There has also been some
progress to identify the typical companions of the
BSs. \citet{Geller2011} used a statistical argument to point out a
particular companion type, that with the mass of 0.5 $M_{\odot}$,
strongly implicating a WD (and therefore a mass transfer from a giant
origin). Most recently, WDs were indeed confirmed to accompany BSs in
several star clusters via detection of UV excess
\citep[][]{Gosnell2014,Sidhu2019,Sahu2019}. Taken at face value, our
measurements indicate that the contribution of merged stars must be
rather small if the first two formation scenarios are considered. Note
however that in view of our observations mergers are not ruled out if
BS originate in triples.

\begin{figure*}
  \centering
  \includegraphics[width=0.995\textwidth]{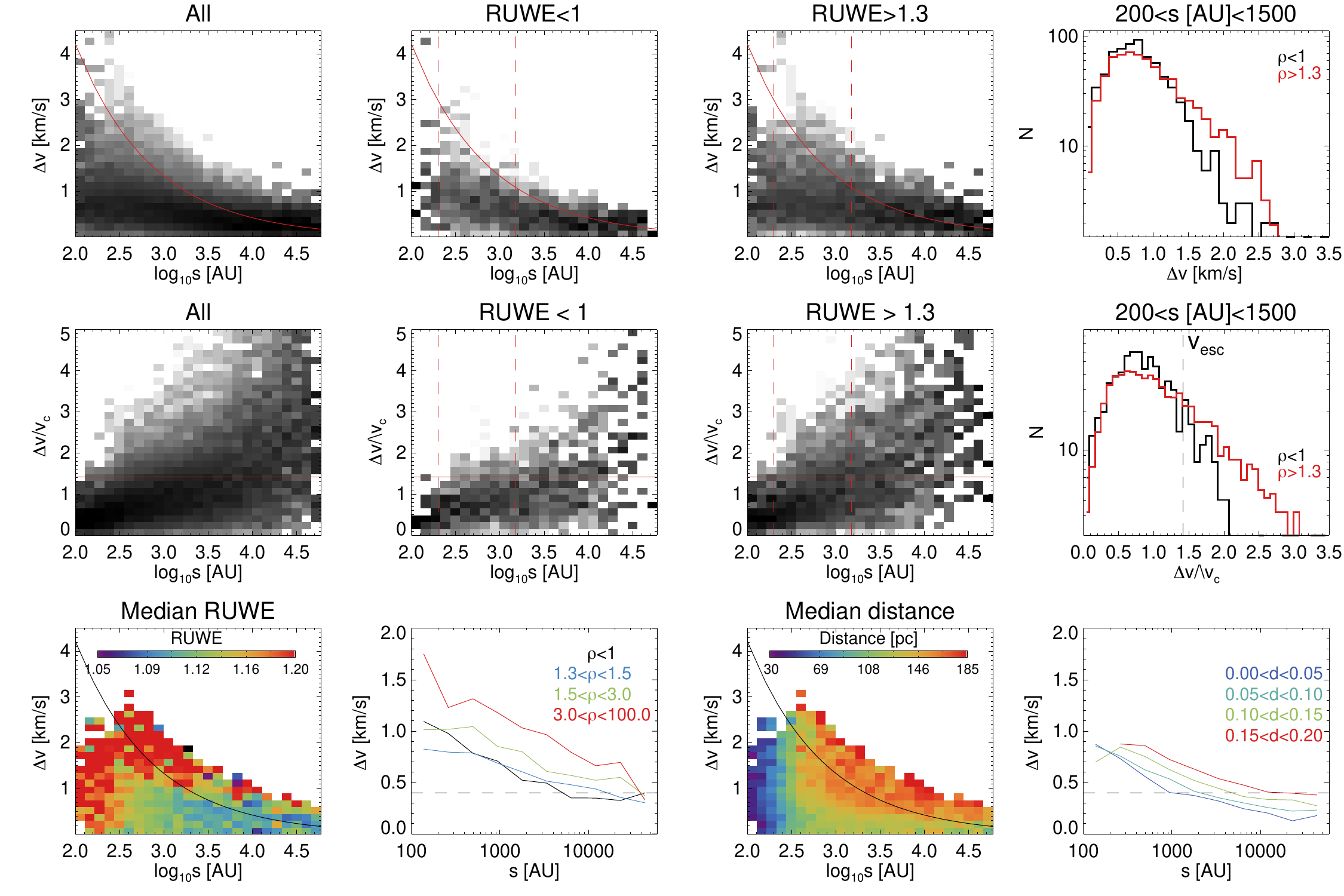}
  \caption[]{$\sim29,500$ wide binaries from the catalogue of
    \citet{ER2018}. Only systems satisfying the criteria described in
    Section~\ref{sec:wide} are shown. {\it Top row:} Relative
    projected velocity (km/s) of the pair as a function of separation
    (AU). {\it 1st panel:} All systems. {\it 2nd panel:} Systems with
    RUWE$<1$. {\it 3rd panel:} Systems with RUWE$>1.3$. Here RUWE is
    the highest RUWE value in the pair. Note that binary systems with
    at least one star showing evidence for photocentre wobble exhibit
    higher relative velocities. {\it 4th panel:} Distributions of
    relative tangential velocities for binary systems with separation
    $200<s ({\rm AU}) < 1500$. {\it Middle row:} Same as top but for
    relative velocity normalised by the circular velocity of the pair
    $v_c$. Note that systems with low RUWE rarely exceed the estimated
    escape velocity $\sqrt{2}v_c$, while those binaries with suspected
    photocentre perturbation clearly do. {\it Bottom row:} RUWE and
    distance as a function of relative velocity and separation. {\it
      1st panel:} Relative velocity as a function of separation
    colour-coded by median RUWE value. Note the RUWE excess at low
    separations and high relative velocities. {\it 2nd panel:}
    Relative velocity trends for systems in four bins of RUWE. {\it
      3rd panel:} Median distance. Note a clear trend of increasing
    distance as a function of separation. {\it 4th panel:} Relative
    velocity trends for systems in four distance bins.}
   \label{fig:wide}
\end{figure*}

\subsection{Position on HB, mass loss and binaries}

Figure~\ref{fig:hr_da} indicates a surprisingly high binary fraction
for stars on the HB. To verify whether this could possibly be due to
an artefact we conduct the following simple test. We select candidate
BHB stars from the sample of objects satisfying the criteria listed in
Equation~(\ref{eq:hrsel}), a cut on tangential velocity $V_T>130$
kms$^{-1}$ and a colour-magnitude selection shown in the left panel of
Figure~\ref{fig:bhb}. For each of the 54 BHB candidates, we identify
30 clones, i.e. stars matching the BHBs in (uncorrected for dust)
$BP-RP$ colour and magnitude $G$. As demonstrated in the right panel
of Figure~\ref{fig:bhb}, the RUWE distribution of the comparison
(clone) sample peaks at $\rho\approx1$ while that of the BHB
candidates is shifted towards higher RUWE, with its peak located at
$\rho\approx1.1$. From this test we conclude that the detected
increase in binarity for the BHB stars is unlikely to have been caused
by the {\it Gaia} systematics.

Low-mass stars (those with mass $<2.5 M_{\odot}$) are expected to shed
significant portions of their outer envelopes as red giants (RGs) in
order to touch down on the long and narrow Horizontal Branch. There
they evolve by burning helium in the core surrounded by a hydrogen
shell, and will slide along the HB left and right before ascending the
Asymptotic Giant Branch (AGB) and disappearing from our view as white
dwarfs (WDs) \citep[see e.g.][and references
  therein]{Catelan2007}. The RGB mass loss controls the exact
placement of a star on the HB and therefore governs its subsequent
evolution. However, no solid theoretical explanation of the mass loss
exists to date. Instead, current stellar evolution models rely on
simple mass loss parameterizations \citep[e.g.][]{Reimers1975,
  SC2005}. This lacuna in the otherwise physically-motivated stellar
evolution theory is also notoriously difficult to make good through
direct observations
\citep[e.g.][]{Origlia2007,Groenewegen2012,Origlia2014}. One of the
best known applications of the RGB mass loss ansatz is the
interpretation of the observed diversity of the globular cluster (GC)
HBs. The HB morphology (its temperature profile) is explained by the
degree to which the helium core is exposed. The GC data indicate that
the primary factor responsible for the HB diversity is the cluster's
metallicity, while the age and the He abundance may act as the second
and third parameters \citep[][]{Gratton2010}. Curiously, using the
same GC HB data to calibrate the mass-loss laws mentioned above,
\citet{McDonald2015} find little dependence on metallicity but a
relatively high mass-loss rate. This result is contradicted by
\citet{Heyl2015} who show that at least in the case of 47 Tuc, the RGB
mass loss is minimal, adding to the growing body of evidence that the
RGB mass loss rates may be significantly overestimated
\citep[e.g.][]{Meszaros2009}. In the absence of a working theory of
RGB mass loss, other scenarios facilitating mass removal have been
suggested. For example, stellar fly-bys and binary interactions can
provide pathways to transfer mass away from the RGB or enhance its
wind \citep[][]{Tout1988,Fusi1993,Buonanno1997,Lei2013,Pasquato2014}.

\begin{figure*}
  \centering
  \includegraphics[width=0.995\textwidth]{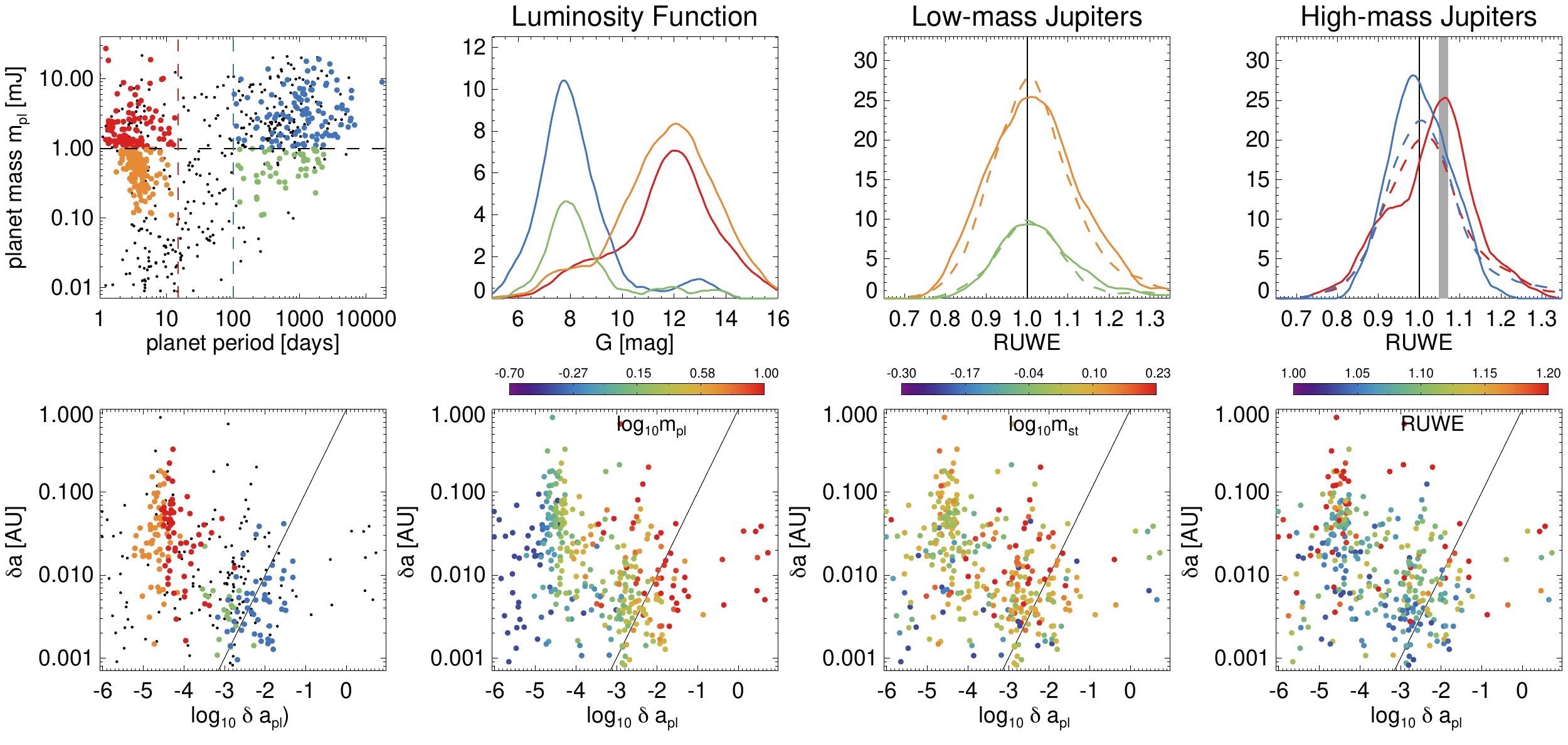}
  \caption[]{Some $\sim$2000 previously known exoplanet hosts as seen
    by Gaia (please see the main text for the sample definition). {\it
      Top row, 1st panel:} Exoplanet mass as a function of its orbital
    period. Colours illustrate the selection of hosts of high-mass
    (red) and low-mass (orange) hot jupiters together with those of
    high-mass (blue) and low-mass (green) outer jupiters. {\it Top
      row, 2nd panel:} Apparent magnitude distributions for the four
    groups selected as shown in the 1st panel. Note that the outer
    jupiter hosts are on average much brighter than the hot jupiter
    hosts. This is the consequence of the detection technique
    selection bias. {\it Top row, 3rd panel:} RUWE distributions for
    the low-mass jupiters (solid lines) together with their comparison
    samples (dotted lines). {\it Top row, 4th panel:} Same as previous
    panel but for the high mass jupiters. Note that the peak of the
    RUWE distribution for the high-mass hot jupiters is shifted to a
    value higher than $\rho=1$, while the peak of the RUWE
    distribution of the corresponding comparison population remains at
    1. {\it Bottom row:} Photocentre wobble $\delta a$ as a function
    of the predicted source displacement if it was caused by the
    catalogued exoplanet $\delta a_{\rm pl}=[m_{\rm pl}/(m_{\rm
        st}+m_{\rm pl})]a$. {\it 1st:} Same colour-coding as in the
    1st panel of the top row. {\it 2nd:} Points are colour-coded
    according to the planet mass. {\it 3rd:} Colour-coding according
    to the host mass. {\it 4th:} Colour-coding reflects the host's
    RUWE value. All of the above distributions are obtained with the
    optimal kernel size KDE (Epanechnikov kernal).}
   \label{fig:jupiters}
\end{figure*}

\subsection{Wide binaries and hierarchical triples}
\label{sec:wide}

Recently, it was suggested that the {\it Gaia} kinematics of wide
binary systems can be used as a gravity test, probing the regime of
weak accelerations \citep[][]{Pittordis2018, Hernandez2019}. By
examining relative velocities as a function of the binary separation,
we find a substantial number of systems for which the velocity
difference exceeds that of the predicted escape speed
\citep[see][]{Pittordis2019}. However, this high velocity tail does
not necessarily require a modification of our gravity theory. As
\citet{Clarke2019} shows, high relative velocities at large
separations can be explained instead by a contribution from
hierarchical triples where the smallest separation binary sub-system
is unresolved by {\it Gaia}. For such an unresolved binary, if the
luminosity ratio is not unity, the photocentre exhibits an additional
excursion due to the binary's orbital motion, biasing the relative
velocity in the wide binary (in reality, a hierarchical triple). Below,
we provide an empirical test of this hypothesis.

Figure~\ref{fig:wide} presents the distribution of projected relative
velocities (computed using the proper motion of the binary components)
as a function of separation for about $29,500$ wide binary systems in
the catalogue of \citet{ER2018}. Note that only MS-MS systems are used
and we require that both components are bright, $G<17$, and suffer
little dust extinction $E(B-V)<0.3$. The first (leftmost) panel in the
top row shows the whole sample, while the second and third panels
display low- and high-RUWE sub-samples correspondingly. As a
reference, the red solid curve gives the escape velocity for a 1
$M_{\odot}$ system. Note that for each binary, the highest of the two
individual RUWE values is chosen. Comparing the second and third
panels, it is immediately clear that high-RUWE stars achieve higher
relative velocities at given separation. This is further illustrated
in the fourth panel of the top row, using one-dimensional velocity
distributions for a range of separations between 200 and 1500 AU
(marked by the red vertical dashed lines in the second and third
panels). While the relative velocity distribution of the low-RUWE
stars begins to drop quickly around 1.5 km s$^{-1}$, the high-RUWE
histogram extends out to about $3$ km s$^{-1}$. Using equation 3 of
\citet[][]{Pittordis2019}, we can compute the masses of the stars in a
binary and thus the corresponding circular orbit velocity $v_c$ and
the associated escape velocity $\sqrt{2}v_c$. The middle row of
Figure~\ref{fig:wide} gives the behaviour of the relative velocity
normalised by $v_c$. Here, the horizontal red line marks the escape
velocity. By comparing the behaviour of the whole sample and the low-
and high-RUWE subsets, we see that the bulk of the systems with
relative velocity exceeding $\sqrt{2}v_c$ are those showing strong
evidence for an additional photocentre perturbation, possibly caused
by an unresolved companion, in line with the hypothesis of
\citet{Clarke2019}. Note that not all wide binaries with the relative
velocity excess have high RUWE. Instead in these systems the period of
the unresolved binary can be larger than {\it Gaia}'s baseline,
thus yielding a well-behaved astrometric fit but noticeable proper
motion anomaly (see also Penoyre et al, submitted).

The first panel in the bottom row of Figure~\ref{fig:wide} gives the
distribution of RUWE in the plane of relative velocity and
separation. Here, several trends are immediately noticeable. First,
RUWE grows with increasing relative velocity, or rather, higher
relative velocities can be achieved by unresolved small-separation
binaries due to an additional photocentre wobble. Secondly, for
separations $s<200$ AU, most of the stars have elevated RUWE. Note
that the catalogue of \citet{ER2018} is limited to 200 pc from the
Sun. Therefore, stars at these separations are less than
1$^{\prime\prime}$ apart on the sky. While {\it Gaia} can resolve most
of these systems, the quality of the astrometric fit would most likely
be affected by the presence of a bright neighbour in the close
proximity. Finally, the median RUWE appears to decrease with growing
separation. To understand this, it is instructive to look at the map
of the median distance shown in the third panel of the bottom row. The
distance distribution shows two trends. Starting at low separations,
the distance increases with growing $s$. This is understandable
because small separation binaries can only be resolved when they are
nearby. There is an additional trend for $\log_{10}(s/{\rm AU})>3$,
where distance also appears to correlate with relative velocity. Many
of the systems above the fiducial escape curve, that have been
demonstrated above to have high RUWE are also typically the most
distant stars in the sample. We hypothesise that this trend could
possibly be due to the fact that high RUWE stars are unresolved
binaries and thus are intrinsically brighter and therefore detectable
to larger distances. Given these distance gradients, it is now clear
why median RUWE decreases with increasing separation: the amplitude of
the photocentre perturbation drops with distance as shown by
Equations~\ref{eq:dtheta} and \ref{eq:da}. Additionally, at larger
distances, the contribution of unresolved binaries with proper motion
anomaly (and periods larger than 22 months) likely becomes more
important.

Based on the analysis presented in Section~\ref{sec:frachrd}, we can
estimate the fraction of hierarchical triples amongst the wide
binaries in the sample considered. First, RUWE values are drawn for
pairs of stars from the single-source RUWE distribution following the
prescription outlined in Section~\ref{sec:ruwe}. Next the maximal RUWE
for the pair is calculated to mimic the procedure we have applied to
the wide-binary sample. The centroiding errors and distances are
propagated to produce a distribution of $\delta a$ for pairs of
sources without statistically significant RUWEs. We subtract the
resulting mock single-source $\delta a$ distribution from the measured
$\delta a$ distribution and sum up the positive differences. In doing
so, we assume that the bias owing to the binary systems being resolved
at low distances and the increasing contribution of the amplified
background cancel each other. We have checked that this indeed appears
to be the case, because our triple fraction estimate changes very
little with the distance cut applied. We report the fraction estimated
for systems between 30 and 150 pc (to avoid the extremes of the biases
mentioned).  The sample average fraction of triples for this sample is
$\sim40\%$, which agrees well with the estimates in the literature
\citep[see e.g.][]{Riddle2015} and is only slightly lower than the
fraction assumed in the calculation of \citet{Clarke2019}. As
explained above, binaries with periods longer than 22 months will have
little RUWE excess and thus our triple fraction estimate might well be
a lower bound on the true value. Going back to the original idea of
testing the theory of gravity in the regime of weak interactions with
wide binaries, it is now clear that stellar multiplicity has to be
carefully taken into account, as already indicated by
\citet{Clarke2019}.

\subsection{Hot Jupiter hosts}

Hot jupiters are difficult to produce at their observed separations
from host stars so they are hypothesised to have migrated from larger
distances either through angular momentum loss inside a remnant
circumstellar disc \citep[e.g.][]{Goldreich1980,Lin1986} or via
interactions with other companions of their hosts, perhaps through the
Kozai-Lidov mechanism \citep[e.g.][]{Wu2003,Fabrycky2007}. The former
pathway leads to a predominantly circular planetary orbit which is
well aligned with the star's spin axis, while the latter yields a
preference for planets with misaligned and eccentric orbits. To test
the importance of multi-body interactions for hot jupiter orbital
evolution, several observational experiments have been carried out
recently
\citep[e.g.][]{Knutson2014,Wang2015,Wollert2015,Wollerta2015,Ngo2015,Piskorz2015,Ngo2016,EvansLucky2016}. In
particular, \citet{Ngo2016} and \citet{Fontanive2019} find a
statistically significant excess of wide-separation (typically beyond
tens of AU) companions to hot jupiters compared to the field
stars. They conclude that it is unclear how exactly such wide
companions can affect the planet's formation and/or
evolution. However, most recently, \citet{Moe2019} re-assessed
selection biases of the above surveys for planetary host companions
and concluded that hot jupiter hosts showed no statistically
significant preference for wide companions.

Figure~\ref{fig:jupiters} presents the properties of exoplanet hosts
as reported by the NASA Exoplanet Archive on 6 September 2018 and
cross-matched with the {\it Gaia} DR2 using 1$^{\prime \prime}$
aperture\footnote{see \url{https://gaia-kepler.fun} for details}. The
first (left) panel in the top row of the Figure shows the exoplanet
mass (in Jupiter masses) as a function of its orbital period. The host
stars should have an apparent magnitude in $5<G<16$ and have no
neighbouring sources within 2$^{\prime \prime}$ as detected by {\it
  Gaia}. Also, the stars should be within 1 kpc from the Sun, have
$\varpi/\sigma_{\varpi}>7$ and lie on the MS, $M_G>2.7$. Finally,
hosts need to have at least 10 comparison stars in a
0.125$\times$0.125 mag pixel on the apparent (BP-RP, G)
colour-magnitude diagram. The combination of the above selection
criteria results in a sample of 1938 exo-planetary hosts out of the
3,666 in the original sample. From this restricted dataset, we select
four groups, low- and high- mass hot jupiters (orange and red) with
periods less than 15 days as well as low- and high-mass outer
jupiters, those with periods greater than 100 days (green and blue).
Of these, 111 are high-mass hot jupiters and 147 are low-mass hot
jupiters. There are also 109 high-mass outer jupiters and 44 low-mass
outer jupiters. The second panel in the top row of
Figure~\ref{fig:jupiters} displays the apparent magnitude
distributions for the four selected exoplanet host groups. These
luminosity functions reveal a strong selection bias due to a
particular detection method involved. The absolute majority of outer
jupiters have so far been detected using radial velocity
identification and thus their host stars are significantly brighter
than those discovered by the transit method.

\begin{figure}
  \centering
  \includegraphics[width=0.49\textwidth]{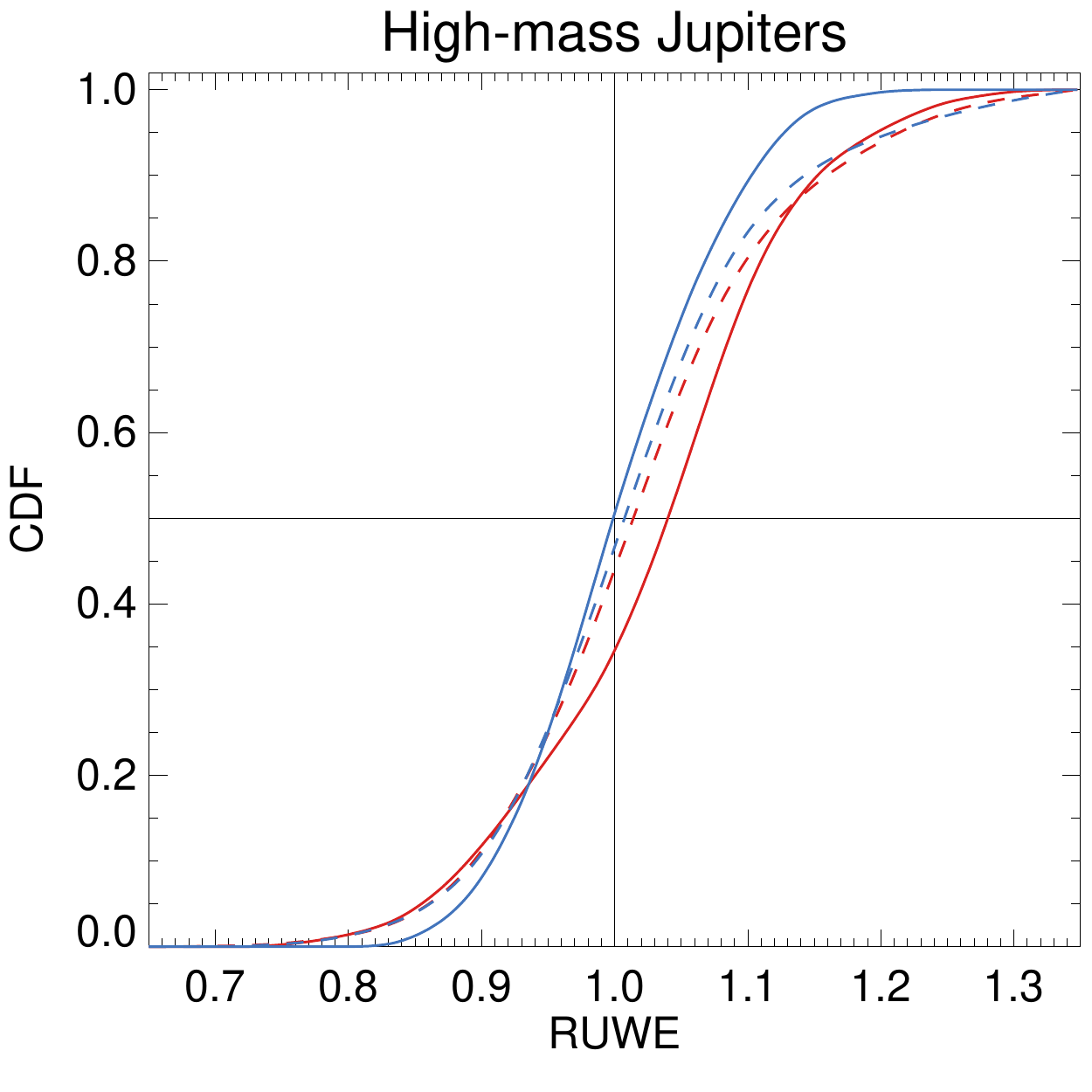}
  \caption[]{Cumulative distribution functions obtained with optimal
    kernel size KDE (Epanechnikov kernel) for the high-mass hot
    jupiters (solid red), high-mass outer jupiters (solid blue) and
    their CMD clones (dashed).}
   \label{fig:jupcdf}
\end{figure}

The third and fourth panels in the top row of
Figure~\ref{fig:jupiters} give the RUWE distributions for the four
selected exoplanet host groups together with the corresponding
distributions for their comparison samples (dashed lines). The
comparison samples are constructed as follows. For each of the
exoplanet hosts, we identify 10 {\it Gaia} DR2 stars with matching
colors and apparent magnitude. The low-mass jupiter hosts (for the
planets both near and far from the host) exhibit RUWE distributions
indistinguishable form those of their comparison samples (see the
third panel in the top row of the Figure). The high-mass samples, in
particular that of the hot jupiters, show small but clear deviations
from their comparison sets that are worth discussing. As revealed by
the blue line, the high-mass outer jupiter hosts have slightly lower
RUWE values overall, and are missing the low-amplitude tail of high
RUWEs. This could be due to the fact that only the most well-behaved
stars are used for the radial velocity exoplanet detection. On the
other hand, the mode of the high-mass hot jupiter distribution is
shifted off $\rho=1$ towards larger $\chi_{\nu}^2$ values, i.e. the
bulk of the hot jupiter hosts show a small but systematic RUWE
excess. The simplest interpretation is that these systems harbour an
additional, probably low-mass and/or distant, stellar or sub-stellar
component, which causes the photocentre to wobble. A grey vertical band
marks the median RUWE value for 13 hosts with known companions within
1.1$^{\prime\prime}$, namely: HAT-P-14, HAT-P-8, HD 68988, HD 86081,
WASP-36, WASP-3, XO-5, HAT-P-24, HAT-P-30, HAT-P-33, HD 109749,
WASP-48 and WASP-76
\citep[see][]{friends,friends2,friends3,Evans2016,Bryan2016}. Curiously,
the median RUWE for the exoplanet hosts with known low-luminosity and
small-separation companions matches rather well the mode of the
distribution of the entire hot jupiter sub-sample.

The bottom row of Figure~\ref{fig:jupiters} translates RUWE into
$\delta a$, the amplitude of the photocentre perturbation. From the
first (left) panel in the bottom row, it is evident that hot jupiter
hosts attain higher amplitude perturbations compared to those of outer
jupiters. This is perhaps not surprising given that the hot jupiter
hosts are typically much further away, thus the apparent angular
astrometric wobble is scaled up by a higher typical distance. The
horizontal axis in all four panels of the bottom row is the amplitude
of the photocentre perturbation $\delta a_{\rm pl}$ if it were caused
by the planet itself \citep[see e.g.][]{Perryman2014}. It is
immediately clear that $\delta a$ and $\delta a_{\rm pl}$ are
completely uncorrelated. Indeed it is the outer jupiters that should
cause a larger photocentre perturbation, because they span a similar
mass range but are located at significantly larger distances from the
host. Thus, we conclude that the photocentre perturbation as revealed
by RUWE is not induced by the known planets. The next three panels in
the bottom row give the same distribution as that shown in the first
panel but color-coded by the planet mass (second panel), the host mass
(the third panel) and RUWE (the fourth panel). Apart from a modest and
expected sorting with the planet mass, no other correlations are
apparent.

Figure~\ref{fig:jupcdf} compares the cumulative distribution functions
of the high-mass hot and outer jupiters (solid lines) and their clone
samples (dashed lines). Running a Kolmogorov-Smirnov test on the
high-mass hot jupiters and their clones, we obtain the probability of
0.03 that the two samples come from the same distribution. This
probability goes down to 0.009 if we limit the host star distances to
0.5 kpc (and the number of hosts to 90). This indicates that there may
indeed be a modest amount of statistical significance present in the
RUWE excess of the high-mass hot jupiters. High-mass outer jupiters
and their clones have a probability of 0.003 to have come from the
same distribution, which also supports the idea that the differences we
saw, i.e. evidence for selection biases favouring well-behaved
stars, may be genuine \citep[see e.g. discussion in][]{Moe2019}. For
the low-mass systems, the KS test supports the hypothesis that the
RUWE values for the planet hosts and their clones come from the same
distribution, for the hot low-mass jupiters with a probability of 0.38
and for the low-mass outer jupiters with a probability of 0.27. These
numbers only increase if a distance cut of 0.5 kpc is applied.

\section{Conclusions}

We have demonstrated that for stars with unresolved companions, the
reduced $\chi^2$ of a single-source astrometric fit provided as part
of {\it Gaia} DR2 can be used to gauge the amplitude of the star's
photocentre perturbation induced by the system's orbital motion. In
practice, we work with the renormalized unit weight error (RUWE) or $\rho$, 
the square root of the reduced $\chi^2$ rescaled to peak at $\rho=1$ across 
the entire colour-magnitude space observed by {\it Gaia}.

Using a sample of known spectroscopic binaries, we have shown that the
amplitude of the angular centroid wobble drops inversely proportional
to the star's distance, as expected. The photocentre displacement
scaled by the source distance, $\delta a$, increases with the binary's
separation and mass. Predictably, {\it Gaia}'s sensitivity to
astrometric binaries is a function of distance as well as mass and
luminosity ratios. We estimate that for systems within 1-2 kpc from
the Sun, systems with semi-major axis size between 0.1 and 10 AU can
be detected. Smaller separation binaries, i.e. those with $\delta
a/{\rm AU}<0.1$ can still be identified if they are nearby. Wider
binaries, corresponding to periods longer than several years do not
produce a significant RUWE excess because the centroid displacement -
as observed by {\it Gaia} - is quasi-linear and is absorbed into the
proper motion. Nonetheless, {\it Gaia} can pick these systems just as
well, not with RUWE excess, but using the so-called proper motion
anomaly instead (see Penoyre et al. for details)

We have also identified situations when the object's RUWE can increase
due to factors not related to the binary orbital motion. For example,
a semi-resolved double star (a genuine binary or a chance alignment)
identified as a single object by {\it Gaia} can have an excess RUWE
due to the mismatch between the model PSF and the observed
image. Additionally, variable stars with large amplitudes of flux
change can accumulate significant RUWE surplus. This happens by
design, because the normalizing coefficient for RUWE is calculated
under the assumption of constant apparent magnitude.

Taking advantage of {\it Gaia}'s unprecedented sensitivity, we have
measured stellar binary fraction for a wide spectrum of stellar
populations across the Hertzsprung--Russell diagram. In the range of
semi-major axes accessible to {\it Gaia}, the binary fraction changes
dramatically as a function of star's mass and its evolutionary
phase. On the Main Sequence, binarity is at its highest for the
youngest (and thus most massive) stars and drops steadily as one moves
to lower mass stars, eventually reaching the level of only few percent
at the bottom of the MS. We find that local Blue Stragglers and Blue
Horziontal Branch stars both show high levels of binarity. The lowest
incidence is observed for double white dwarfs, of which only 1\% are
detected to reside in unresolved binaries. Note this number needs to
be corrected for the imposed selection effects and the overall double
WD fraction is higher but unlikely to exceed 10\%.

Analysing the astrometric properties of wide binary companions, we
have measured a high occurrence rate of hierarchical triples. The
inner binary sub-system is typically unresolved by {\it Gaia} but can
be detected when it exhibits RUWE excess. Such unresolved inner
binaries yield an additional contribution to the relative outer binary
velocity \citep[in agreement with][]{Clarke2019}. Our estimate of
$\sim40\%$ triple incidence amongst wide binaries is only a lower
limit. This is because a single-source fit works well for systems with
periods longer than the {\it Gaia} DR2 temporal baseline (see Penoyre
et al.).

Finally, we have explored the astrometric behaviour of exo-planetary
hosts. Typically, only low-mass stars with massive planets on wide
orbits have a chance of being perturbed enough to be detected by {\it
  Gaia}. It is therefore surprising to see a hint of an excess of hot
jupiter hosts with small but measureable centroid perturbations. In
these systems, the planets are too close to their hosts to perturb the
star's motion. We conclude therefore that, if real, this detection can
be interpreted as a possible evidence for the presence of additional
low-mass companions to massive hot jupiter hosts.

\section*{Acknowledgments}

The authors are grateful to Simon Jeffery, Floor van Leeuwen, Mark
Wyatt, Annelies Mortier, Nikku Madhusudhan, Ingrid Pelisoli, Silvia
Toonen and members of the Cambridge Streams Club for many illuminating
discussions that helped to improve this manuscript. This research made
use of data from the European Space Agency mission Gaia
(http://www.cosmos.esa.int/gaia), processed by the Gaia Data
Processing and Analysis Consortium (DPAC,
http://www.cosmos.esa.int/web/gaia/dpac/consortium). Funding for the
DPAC has been provided by national institutions, in particular the
institutions participating in the Gaia Multilateral Agreement. This
paper made used of the Whole Sky Database (wsdb) created by Sergey
Koposov and maintained at the Institute of Astronomy, Cambridge with
financial support from the Science \& Technology Facilities Council
(STFC) and the European Research Council (ERC). This work made use of
the gaia-kepler.fun crossmatch database created by Megan Bedell. SK
acknowledges the support by NSF grants AST-1813881, AST-1909584 and
Heising-Simons foundation grant 2018-1030.

\appendix

\section{RUWE distribution across the CMD, connection to AEN}

\begin{figure*}
  \centering \includegraphics[width=0.99\textwidth]{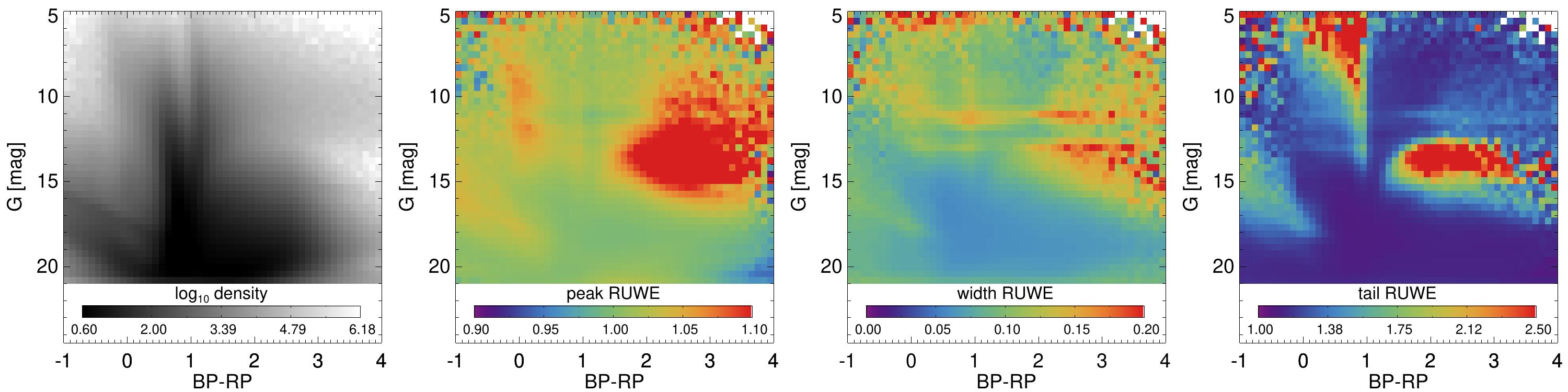}
  \caption[]{ {\it 1st panel:} Logarithm of the density of sources
    with $|b|>15^{\circ}$ as a function of extinction-corrected color
    \bprp\ and apparent magnitude $G$. The pixel size is
    0.1$\times$0.59 mag, the total numbers of objects
    $\sim2.8\times10^8$. {\it 2nd panel:} Peak of the RUWE
    distribution as captured by the 41st percentile \citep[see][for
      details]{RUWE}.  {\it 3rd panel:} The width of the RUWE
    distribution (standard deviation of the RUWE distribution). {\it
      4th panel:} The tail of the RUWE distribution as tracked by the
    90th percentile.}
   \label{fig:cmd_ruwe}
\end{figure*}

Figure~\ref{fig:cmd_ruwe} shows the properties of the RUWE
distribution as a function of the position on the Color-Magnitude
Diagram spanned by extinction-corrected color \bprp\ and apparent
magnitude $G$ for $\sim2.8\times10^8$ sources with
$|b|>15^{\circ}$. The first panel of the Figure shows the density of
the sources in the CMD space. Even though the distribution is
convolved with the line-of-sight density profile, many familiar
features, such as MS, RGB, WD etc, can be discerned. The second panel
of the Figure gives the map of the RUWE peak - as traced by the 41st
percentile \citep[see][]{RUWE} - which appears to stay around 1 in all
of the well-populated regions of the CMD. Note a slight positive shift
of the peak in the blue part of the CMD, i.e. for BP-RP$<0.5$. Note
additionally, a region in the red part of the CMD where the peak is
shifted to at least 1.1 or slightly higher. This part of the space
corresponds to a poorly populated region, in between the red giants
and red dwarfs. The width of the RUWE distribution is given in the
third panel of the Figure. It appears that the width does not vary
dramatically as a function of color and magnitude, with a typical
value of $<0.1$ and a pattern of deviation which appears to track that
shown in the previous panel. Finally, the behaviour of the tail of the
RUWE distribution (captured by the 90th percentile) can be seen in the
fourth panel of the Figure.

\begin{figure}
  \centering \includegraphics[width=0.49\textwidth]{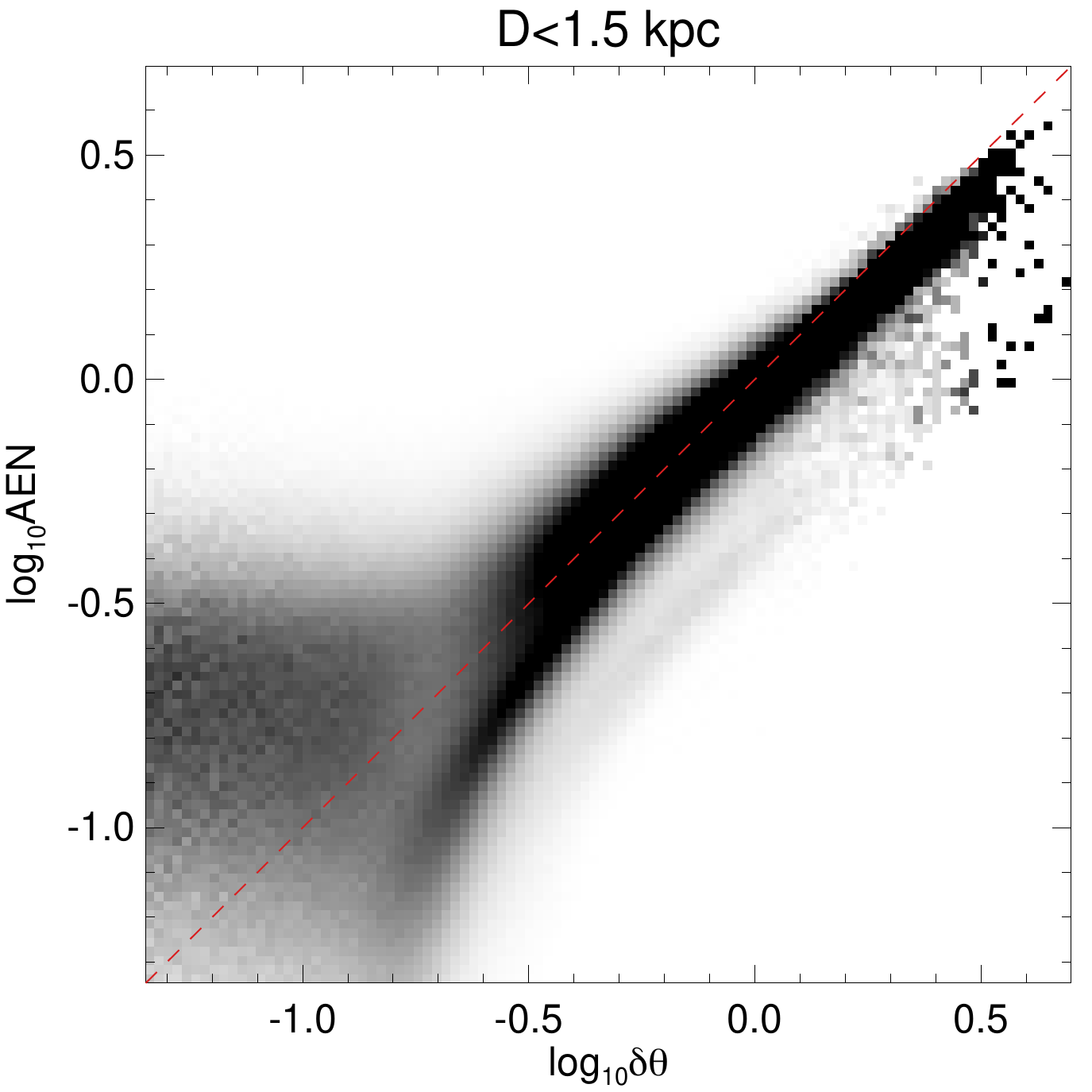}
  \caption[]{Correspondence between astrometric excess noise (AEN) and
    the angular centroid shift amplitude $\delta\theta$ calculated
    using equations~\ref{eq:dtheta} and \ref{eq:da}.}
   \label{fig:aen}
\end{figure}

Figure~\ref{fig:aen} demonstrates that, as expected, the angular
centroid shift amplitude $\delta\theta$ (calculated using
equations~\ref{eq:dtheta} and \ref{eq:da}) correlates tightly with
the astrometric excess noise (AEN).

\section{$\delta_{\rm ql}$ for model stellar populations}
\label{sec:dql}

\begin{figure*}
  \centering
  \includegraphics[width=0.99\textwidth]{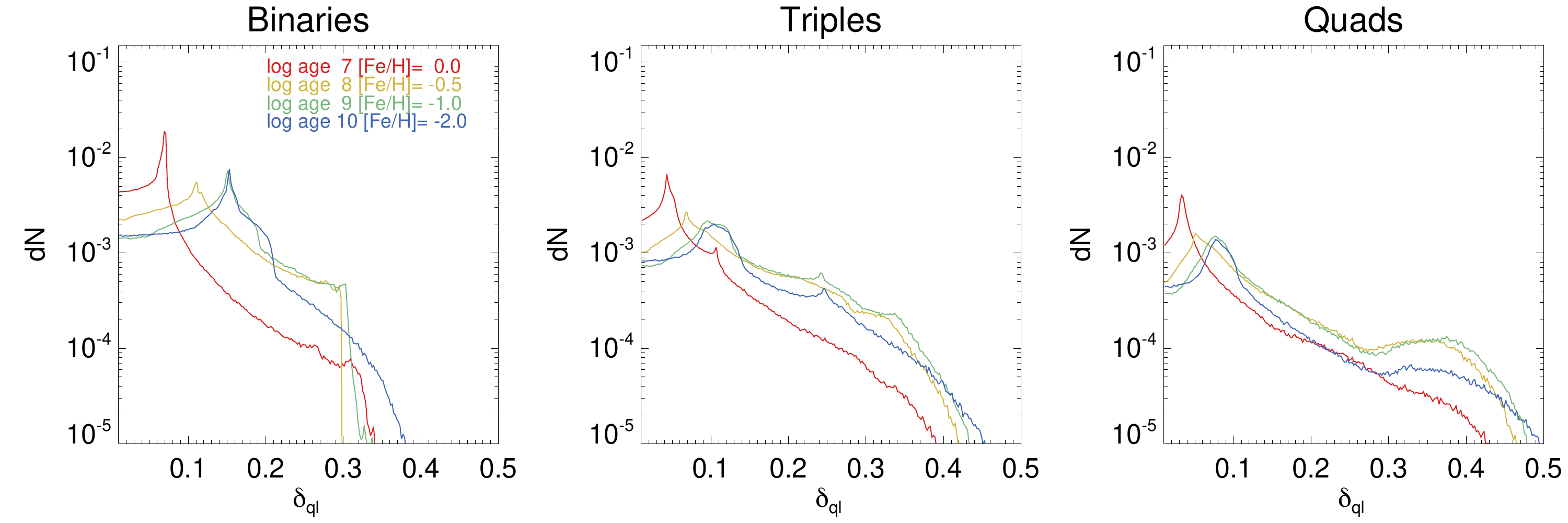}
  \caption[]{Astrometric shift scaling factor $\delta_{ql}$ for
    different stellar population in binaries (top), triples (center)
    and quadruples (bottom). To produce $\delta_{ql}$ distribution we
    use PARSEC isochrones with ages and metallicities shown in the top
    right corner of the top panel. Stellar masses and luminosities are
    drawn from the corresponding mass function. Triples (quadruples)
    are assumed to be in a hierarchical configuration equivalent to a
    binary where one component is a binary (triple) itself. Note that
    shapes of the distributions depend on the properties of the
    stellar population considered, e.g. for young metal-rich stars
    (red curve), a typical values of $\delta_{ql}\sim0.05$. Higher
    multiplicity systems can achieve higher values of $\delta_{ql}$,
    typically by a factor of 2 to 4.}
   \label{fig:delta}
\end{figure*}

Figure~\ref{fig:delta} gives distribution of $\delta_{ql}$ values for
four PARSEC isochrones with different ages and metallicites for
binaries (top), triples (center) and quadruples (bottom, these are
chosen to be hierarchical systems equivalent to a binary in which one
companion is a combination of three stars). As is clear from the
Figure, $\delta_{ql}$ varies depending on the stellar population and
typically ranges from 0.01 to 0.2. Note that adding a compact (dark)
remnant such as white dwarf, neutron star or a black hole immediately
pushed $\delta_{ql}$ high, i.e. to values close to 1 (for any of the
isochrones considered). Systems of higher multiplicity (such as
triples and quadruples) exhibit extended tails that stretch to higher
values of $\delta_{ql}$, typically increasing it by a factor of
2. Higher multiples, however, can not yield $\delta a$ values higher
by an order of magnitude. We conclude therefore that the while many of
the stars located in the second, low-amplitude bump of the $\delta a$
distribution shown in Figure~\ref{fig:error} may well be in triples
etc, the highest $\delta a$ systems, i.e. those around $\delta a
\sim1$, require a dark (faint) companion, most likely a white dwarf
(assuming similar distribution of semi-major axes).

\bibliography{references}

\label{lastpage}

\end{document}